\documentclass[aps,prb,twocolumn,floats,showpacs,superscriptaddress]{revtex4}

\usepackage{epsfig}
\usepackage{color}
\usepackage{amsmath}

\usepackage{amssymb}

\usepackage{bm}

\def\be{\begin{equation}}
\def\ee{\end{equation}}
\def\bea{\begin{eqnarray}}
\def\eea{\end{eqnarray}}
\def\nn{\nonumber\\}
\def\eps{\varepsilon}

\def\th{\theta}
\begin{document}

\title{Dynamical Correlations of the Spin-$\frac{1}{2}$ Heisenberg XXZ
Chain in a Staggered Field}
\author{Igor Kuzmenko}
\affiliation{The Rudolf Peierls Centre for Theoretical Physics,
University of Oxford, 1 Keble Road, Oxford OX1 3NP, UK}
\affiliation{Department of Physics, Lancaster University,
Lancaster, LA1 4YB, UK}
\author{Fabian H.L. Essler}
\affiliation{The Rudolf Peierls Centre for Theoretical Physics,
University of Oxford, 1 Keble Road, Oxford OX1 3NP, UK}
\date{\today}

\begin{abstract}
We consider the easy-plane anisotropic spin-$\frac{1}{2}$
Heisenberg chain in combined uniform longitudinal and transverse
staggered magnetic fields. The low-energy limit of his model is
described by the sine-Gordon quantum field theory. Using methods
of integrable quantum field theory we determine the various
components of the dynamical structure factor. To do so, we derive
explicit expressions for all matrix elements of the low-energy
projections of the spin operators involving at most two particles.
We discuss applications of our results to experiments on
one-dimensional quantum magnets.
\end{abstract}


\maketitle

\section{Introduction}
The field-induced gap problem in anisotropic quasi one dimensional
spin-$\frac{1}{2}$ Heisenberg antiferromagnets has attracted much
experimental \cite{magn,dender,Asano00,Ajiro00,Asano02,Nojiri06,Kenzelmann,
feyer,Wolter03b,Zvyagin04,Zvyagin05,Wolter05,Oshikawa99,kohgi,chen} and theoretical
\cite{oa97,ET97,Essler99,oa99,oa02,Lou02,Dmitriev02a,Dmitriev02b,EFH03,CEL03,Wolter03b,capraro,zhao03,lou05}
attention in recent years. Two scenarios have been studied in
particular. For isotropic exchange interaction a gap can be
induced by the application of a uniform magnetic field in presence
of a staggered g-tensor and/or a Dzyaloshinskii-Moriya interaction
\cite{oa97}. This is the case for materials such as Copper
Benzoate
\cite{magn,dender,Asano00,Ajiro00,Asano02,Nojiri06}, CDC
[${\rm CuCl_2\cdot 2((CD_3)_2SO)}$]
\cite{Kenzelmann}, Copper-Pyrimidine
\cite{feyer,Wolter03b,Zvyagin04,Zvyagin05,Wolter05} and ${\rm Yb_4As_3}$
\cite{Oshikawa99,kohgi}. Theoretical studies have analyzed the
excitation spectrum \cite{oa97,ET97,Lou02}, the dynamical
structure factor \cite{ET97,EFH03}, the specific heat
\cite{Essler99}, the magnetic susceptibility \cite{oa99,Wolter03b}
and the electron-spin resonance lineshape \cite{oa02}. In the
materials mentioned above application of a uniform magnetic field
$\bm{H}$ induces a staggered field perpendicular to $\bm{H}$. It
is the induced staggered field that leads to a spectral gap. The
staggered field is generated both by a staggered $g$-tensor
\cite{oshima76,oshima78} and a Dzyaloshinskii-Moriya (DM)
interaction. The simplest Hamiltonian describing such
field-induced gap systems is given by \cite{oa97}
\begin{equation}
  {\cal H}=
  J\sum_j
  \bm{S}_j\cdot\bm{S}_{j+1}-
  H\sum_jS_j^z+
  h\sum_j(-1)^jS^x_j,
  \label{hamil}
\end{equation}
where $h=\gamma H$. The constant $\gamma$ is given in terms of the
staggered $g$-tensor \cite{oshima76,oshima78} and the DM
interaction. In critical systems with exchange anisotropy such as
the spin-1/2 Heisenberg XXZ chain a second mechanism for inducing
a gap by application of a uniform magnetic field exists. While
application of a field perpendicular to the easy plane leaves the
system critical, applying a field in the easy plane leads to the
formation of a spectral gap
\cite{Kurmann,KenzelmannPRB65,Dmitriev02a,Dmitriev02b,CEL03,capraro}.

The purpose of the present work is to extend the theoretical
analysis of the staggered field mechanism for generating a
spectral gap to the case of the anisotropic Heisenberg chain,
\begin{eqnarray}
  {\cal H} &=&
  J\sum_{j}
  \Big[
      S^x_jS^x_{j+1}+
      S^y_jS^y_{j+1}+
      \delta
      S^z_jS^z_{j+1}
  \Big]-
  \nonumber \\ && -
  H\sum_{j}S^z_j+
  h\sum_{j}(-1)^jS^x_j.
  \label{H-uni-stagg}
\end{eqnarray}
In what follows we will consider the region $-1<\delta\leq 1$
which corresponds to an ``XY''-like exchange anisotropy. It is
important for our analysis that the staggered field is transverse
to the anisotropy whereas the magnetic field is along the
anisotropy axis. Only in this case does the low-energy limit map
onto an integrable model, the sine-Gordon quantum field theory.

The outline of this paper is as follows: in section
\ref{Sec-model} we construct the continuum limit of the model
(\ref{H-uni-stagg}). In section \ref{Sec-DSF} we derive a spectral
representation of the dynamical structure factor at low energies.
In section \ref{sec:kinematics} we present the calculations for
retarded two-point correlation functions. In section
\ref{sec:results} we present our results for the components of the
dynamical strucure factors. Section \ref{sec:conc} summarizes our
results. The technical aspects of our analysis are summarized in
several appendices: in Appendix \ref{app:BA} we discuss how the
parameters of the low-energy field theory can be determined from
the Bethe ansatz solution of the Heisenberg chain in a magnetic
field. Appendices \ref{Sec-exp-FF-sb} and \ref{Sec-exp-FF-bb}
present results for the form factors of the operators entering the
calculation of the dynamical structure factor.

\section{Continuum Limit}
        \label{Sec-model}
In the limit $|h|\ll H,J$ the staggered field can be taken into
account as a perturbation to the low-energy limit of the XXZ chain
in a magnetic field. It is well-known that the low-energy limit of
the spin-$\frac{1}{2}$ Heisenberg XXZ chain with XY-like
anisotropy $|\delta|<1$ is given by a free bosonic theory
\cite{Luther75,Haldane81a,Haldane81b,Affleck89b}
 \bea
   {\mathcal H}_{h=0}
   &=&
   \frac{v}{16\pi}
   \int dx
   \left[
        (\partial_x\Phi)^2+
        (\partial_x\Theta)^2
   \right],
   \label{GaussianModel}
 \eea
where the field $\Phi(x)$ and its dual field $\Theta(x)$ are
compactified \be \Theta(x)\equiv\Theta(x)+\frac{2\pi}{\beta}\
,\quad \Phi\equiv \Phi+8\pi\beta. \label{compactTheta} \ee The
commutation relation between $\Phi$ and $\Theta$ reads
\begin{equation}
  \Big[
      \Theta(x),\Phi(x')
  \Big]=
  8\pi i\vartheta_H(x-x'),
  \label{Commut-Phi-Theta}
\end{equation}
where $\vartheta_H(x)$ is the Heaviside step function, equal to $0$
for $x<0$, $1$ for $x>0$ and $1/2$ for $x=0$. The parameters $v$,
$\beta$ and $k_F$ (see below) in the low-energy theory can be
calculated directly from the Bethe ansatz solution on the XXZ
chain \cite{vladb}. How this is done is briefly reviewed in
Appendix \ref{app:BA}. The results as well as the other parameters 
used are listed in Table \ref{table-m-a-c-b-beta-H} for the anisotropic 
parameter $\delta=0.3$.
In the continuum limit the lattice spin
operators have the following expansions \be
S_j^\alpha=\sum_{a=1}^3 e^{iQ_a^\alpha x} {\cal
  S}_a^\alpha(x)+\ldots\ ,
\label{CLS} \ee
where $x=ja_0$ and $a_0$ is the lattice spacing.
The wavenumbers $Q_a^\alpha$ are \bea
Q^{x}_1&=&-Q^x_3=Q=\frac{\pi}{a_0}-2k_F\ ,\ Q^x_2=\frac{\pi}{a_0}\ ,\\
Q^y_a&=&Q^x_a\ ,\\
Q^z_1&=&0\ ,\ Q^z_2=-Q^z_3=2k_F, \eea where the Fermi momentum is
given by \be k_F=\frac{\pi}{2a_0}(1-2\langle{S^z_j}\rangle). \ee
Here $\langle{S^z_j}\rangle$ is the magnetization per site. The
continuum fields ${\cal S}_a^\alpha$ are given in terms of the
canonical boson $\Phi$ and its dual field $\Theta$ as
\begin{eqnarray}
  {\cal S}_1^x(x) &=&
  \frac{1}{2}
  {\cal A}(H)
  \Big[
      {\cal{O}}^1_{\beta}(x)+
      {\cal{O}}^1_{-\beta}(x)
  \Big],
  \label{Sx-smooth-scalar}
  \\
  {\cal S}_2^x(x) &=&
  c(H)\cos\Big(\beta\Theta(x)\Big),\\
  {\cal S}_3^x(x) &=&  \Bigl({\cal S}_1^x(x)\Bigr)^\dagger,
  \label{Sx-stag-scalar}
\eea \bea
  {\cal S}_1^y(x) &=&
  \frac{1}{2i}
  {\cal A}(H)
  \Big[
      {\cal{O}}^1_{\beta}(x)-
      {\cal{O}}^1_{-\beta}(x)
  \Big],
  \label{Sy-smooth-scalar}
  \\
  {\cal S}_2^y(x) &=&
  c(H)
  \sin
  \Big(
      \beta\Theta(x)
  \Big),
  \label{Sy-stag-scalar}  \\
  {\cal S}_3^y(x) &=&  \Bigl({\cal S}_1^y(x)\Bigr)^\dagger,
\eea \bea
  {\cal S}_1^z(x) &=&
  \frac{a_0}{8\pi\beta}~\partial_x\Phi(x),
  \label{Sz-smooth-scalar}
  \\
  {\cal S}_2^z(x) &=&  \Bigl({\cal S}_3^z(x)\Bigr)^\dagger=
  -\frac{1}{2i}
  a(H)
  {\cal{O}}^1_0(x),
  \label{Sz-stag-scalar}
\end{eqnarray}
where
\begin{equation}
  {\cal{O}}^1_a=
  \exp
  \Big\{
      \frac{i\Phi}{4\beta}+ia\Theta
  \Big\}.
  \label{operator-O}
\end{equation}
We are using normalizations such that
\be
\langle{\cal{O}}^1_a(\tau,x){\cal{O}}^{-1}_{-a}(0,0)\rangle=
\left[\frac{v\tau+ix}{v\tau-ix}\right]^\frac{a}{\beta}\!
\left[\frac{a_0^2}{v^2\tau^2+x^2}\right]^{2a^2+\frac{1}{8\beta^2}}.
\ee
The coefficients $a(H)$, $c(H)$ and ${\cal A}(H)$ have been
determined numerically in Ref. [\onlinecite{HF04}]. The staggered
magnetic field perturbation can be bosonized using
(\ref{Sx-smooth-scalar}) -- (\ref{Sz-stag-scalar}), which leads to
a sine-Gordon model
\begin{eqnarray}
  {\cal H} &=&
  \int{dx}
  \bigg\{
       \frac{v}{16\pi}
       \Big[
           (\partial_x\Phi(x))^2+
           (\partial_x\Theta(x))^2
       \Big]+
  \nonumber \\ && +
       \mu(h,H)\cos(\beta\Theta(x))
  \bigg\},
  \label{H-boson}
\end{eqnarray}
where $\mu(h,H)=hc(H)$. We note that as we have
chosen to bosonize in a finite magnetic field, the cutoff of the
theory is $H$ rather than $J$. However, it is straightforward to
recover the zero field limit (where one bosonizes at $H=0$ and the
cutoff is $J$) in the expressions for the structure factor we give
below.

\subsection{Elementary Excitations}
The sine-Gordon model is integrable and its spectrum and
scattering matrix are known exactly
\cite{Dashen75b,FaddeevKorepin78,Zam77,zamo,Korepin79,Bergknoff}.
In the relevant range of the parameter $\beta$ ($0<\beta<1$) the
spectrum of elementary excitations consists of a
soliton--anti-soliton doublet and several soliton--anti-soliton
bound states called ``breathers''. There are altogether $[1/\xi]$
breathers, where $[x]$ denotes the integer part of $x$ and
$\xi=\frac{\beta^2}{1-\beta^2}$. In order to distinguish the
various single-particle states we introduce labels $s$ and
$\bar{s}$ for solitons and anti-solitons respectively and
$b_1,\ldots,b_{[1/\xi]}$ for breathers. Energy and momentum
carried by the elementary excitations are expressed in terms of
the rapidity $\theta$ as
\begin{equation}
 vP_{\epsilon}=\Delta_{\epsilon} \sinh(\theta),
 \ \ \ \ \
 E_{\epsilon}=\Delta_{\epsilon} \cosh(\theta),
 \label{P-E-epsilon}
\end{equation}
where $\Delta_s=\Delta_{\bar{s}}=\Delta$,
$\Delta_{b_k}\equiv\Delta_k=2\Delta\sin(\frac{\pi\xi k}{2})$. The
soliton gap as a function of parameters $H$ and $h$ is
\cite{Zamolod01,Essler99}
\begin{eqnarray}
  \frac{\Delta}{J} =
  \frac{2v}{Ja_0\sqrt{\pi}}
  \frac{\Gamma(\frac{\xi}{2})}
       {\Gamma(\frac{1+\xi}{2})}
  \Bigg[
       \frac{Ja_0 c(H)\pi}
            {2v}
       \frac{\Gamma(\frac{1}{1+\xi})}
            {\Gamma(\frac{\xi}{1+\xi})}
       \frac{h}{J}
  \Bigg]^{\frac{1+\xi}{2}}.
  \label{delta-strong}
\end{eqnarray}
When $\delta\approx 1$ and the magnetization is small the leading
irrelevant perturbation to the Gaussian model needs to be taken
into account, leading to \cite{oa99}
\begin{eqnarray}
 \frac{\Delta}{J}=
 \bigg(
      \frac{h}{J}
 \bigg)^{\frac{1+\xi}{2}}
 \Bigg[
      B
      \bigg(
           \frac{J}{H}
      \bigg)^{\frac{1}{2}-2\beta^2}
      \Big(2-8\beta^2\Big)^{\frac{1}{4}}
 \Bigg]^{-\frac{1+\xi}{2}},
 \label{Delta}
\end{eqnarray}
where $B=0.422169$.
\subsection{Scattering States}
It is useful to introduce creation and annihilation operators
$A^{\dag}_{\epsilon}(\theta)$ and $A_{\epsilon}(\theta)$ for the
elementary excitations. Here
$\epsilon=s,\bar{s},b_1,\ldots,b_{[1/\xi]}$. The
creation/annihilation operators fulfil the so-called
Faddeev-Zamolodchikov (FZ) algebra
\begin{eqnarray}\label{FZ}
A_{a} (\th_1 ) A_{b} (\th_2 ) &=& S^{a'b'}_{ab} (\th_1-\th_2)
A_{b'} (\th_2) A_{a'} (\th_1);\cr\cr A^\dagger_{a} (\th_1 )
A^\dagger_{b} (\th_2 ) &=& S^{a'b'}_{ab} (\th_1-\th_2)
A^\dagger_{b'} (\th_2) A^\dagger_{a'} (\th_1);\cr\cr A^\dagger_{a}
(\th_1 ) A_{b} (\th_2 ) &=& S^{b'a}_{ba'} (\th_1-\th_2) A_{b'}
(\th_1) A^\dagger_{a'} (\th_1) \nonumber\\ &&+
2\pi\delta_{ab}\delta(\th_1-\th_2).
\end{eqnarray}
Here $S(\theta)$ is the scattering matrix of the sine-Gordon model
\cite{Zam77,zamo,Korepin79}. Multi-particle scattering states of
(anti)solitons and breathers are given in terms of the FZ creation
operators as
\begin{eqnarray}
  | \{\epsilon_n,\theta_n\} \rangle=
  A^{\dag}_{\epsilon_n}(\theta_n)
  \ldots
  A^{\dag}_{\epsilon_1}(\theta_1)
  | 0 \rangle.
  \label{n-particle-state}
\end{eqnarray}
Energy and momentum of these states are
\begin{equation}
 E_{\{n\}}=\sum_{i=1}^{n}E_{\epsilon_i},
 \ \ \ \ \
 P_{\{n\}}=\sum_{i=1}^{n}P_{\epsilon_i}.
 \label{P-E-n-particle}
\end{equation}
The resolution of the identity in the normalization implied by
(\ref{FZ}) reads
\begin{equation}
I= \sum_{n=0}^\infty\sum_{\{\epsilon_j\}}
\int\frac{d\theta_1\ldots d\theta_n}{n!(2\pi)^n} |
\{\epsilon_n,\theta_n\} \rangle
  \langle \{\epsilon_n,\theta_n\}|.
\label{ROI}
\end{equation}

\subsection{Discrete Symmetries}
The Hamiltonian is invariant with respect to charge conjugation
\bea C\Theta C^{-1}&=&-\Theta\ ,\quad C\Phi C^{-1}=-\Phi\ . \eea
The action of the charge conjugation operator $C$ on physical
states follows from
\begin{eqnarray}
  &&
  C|0\rangle=|0\rangle,
  \nonumber \\ &&
  C A_s^{\dag}(\theta) C^{-1}=A_{\bar s}^{\dag}(\theta),
  \label{charge-conjugate}
  \\ &&
  C B_k^{\dag}(\theta) C^{-1}=(-1)^kB_k^{\dag}(\theta).
  \nonumber
\end{eqnarray}
We see that even breathers are invariant under charge conjugation,
while odd breathers change sign. The topological charge
$$
 {\cal Q}=\frac{\beta}{2\pi}
 \int\limits_{-\infty}^{\infty}dx\
 \partial_x\Theta(x),
$$
is a conserved quantity. We will use the conventions in which
soliton/antisoliton and breathers have topological charge $\mp 1$
and zero respectively.

\begin{table}
  \caption{\label{table-m-a-c-b-beta-H}
  Amplitudes ${\cal{A}}$, $a$ and $c$, the dimensionless spin
  velocity $v/Ja_0$, the coupling $\beta$, and the field $H$ as
  functions of the magnetization $m$ for the anisotropic
  parameter $\delta=0.3$. The amplitudes are determined
  in Ref. [\onlinecite{HF04}].}
  \bigskip
\begin{tabular}{|c|c|c|c|c|c|c|}
  \hline
  $m$ & ${\cal{A}}$ & $a$ & $c$ & $v/Ja_0$ & $\beta$ & $H/J$ \\
  \hline
  0.02 & 0.3044 & 0.3953 & 0.5275 & 1.1804 & 0.386192 & 0.09093 \\
  \hline
  0.04 & 0.3065 & 0.3913 & 0.5268 & 1.17114 & 0.385821 & 0.18186 \\
  \hline
  0.06 & 0.3096 & 0.3867 & 0.5256 & 1.15828 & 0.385332 & 0.2598 \\
  \hline
  0.08 & 0.3130 & 0.3817 & 0.5240 & 1.13738 & 0.384573 & 0.35073 \\
  \hline
  0.10 & 0.3173 & 0.3769 & 0.5219 & 1.11423 & 0.383768 & 0.42867 \\
  \hline
  0.12 & 0.3226 & 0.3713 & 0.5194 & 1.08072 & 0.38265 & 0.5196 \\
  \hline
  0.14 & 0.3284 & 0.3661 & 0.5164 & 1.04600 & 0.381535 & 0.59754 \\
  \hline
  0.16 & 0.3354 & 0.3610 & 0.5129 & 1.01244 & 0.380489 & 0.66249 \\
  \hline
  0.18 & 0.3433 & 0.3559 & 0.5088 & 0.966005 & 0.379084 & 0.74043 \\
  \hline
  0.20 & 0.3527 & 0.3508 & 0.5041 & 0.921509 & 0.377775 & 0.80538 \\
  \hline
  0.22 & 0.3642 & 0.3460 & 0.4988 & 0.870927 & 0.376322 & 0.87033 \\
  \hline
  0.24 & 0.3773 & 0.3415 & 0.4929 & 0.813165 & 0.374702 & 0.93528 \\
  \hline
  0.26 & 0.3923 & 0.3371 & 0.4861 & 0.760734 & 0.37326 & 0.98724 \\
  \hline
  0.28 & 0.4102 & 0.3329 & 0.4785 & 0.701482 & 0.371658 & 1.0392 \\
  \hline
  0.30 & 0.4321 & 0.3286 & 0.4699 & 0.651491 & 0.370326 & 1.07817 \\
  \hline
  0.32 & 0.4596 & 0.3253 & 0.4602 & 0.575147 & 0.368318 & 1.13013 \\
  \hline
  0.34 & 0.493 & 0.3222 & 0.4492 & 0.507976 & 0.366572 & 1.16910 \\
  \hline
  0.36 & 0.5342 & 0.3193 & 0.4367 & 0.456492 & 0.365244 & 1.19508 \\
  \hline
  0.38 & 0.588 & 0.3166 & 0.4222 & 0.389204 & 0.363518 & 1.22431 \\
  \hline
  0.40 & 0.664 & 0.3141 & 0.4053 & 0.326360 & 0.361913 & 1.24704 \\
  \hline
  0.42 & 0.769 & 0.3131 & 0.3851 & 0.259866 & 0.360218 & 1.26652 \\
  \hline
  0.44 & 0.934 & 0.3125 & 0.3602 & 0.186561 & 0.358349 & 1.28276 \\
  \hline
  0.46 & 1.214 & 0.3127 & 0.3279 & 0.122936 & 0.356722 & 1.29251 \\
  \hline
  0.48 & 1.89 & 0.3142 & 0.2796 & 0.0448071 & 0.354713 & 1.299 \\
  \hline
\end{tabular}
\end{table}

\section{Dynamical Structure Factor}
         \label{Sec-DSF}
The central object of our study is the inelastic neutron
scattering intensity, which is proportional to \cite{igor}
 \bea
 I(\omega,{\bf{k}}) &\propto&
 \sum_{\alpha,\alpha'}
 \left(
      \delta^{\alpha\alpha'}-\frac{k^{\alpha}k^{\alpha'}}{{\bf{k}}^2}
 \right)
 S^{\alpha\alpha'}(\omega,k).
 \label{intensity}
 \eea
Here $\alpha,\alpha'=x,y,z$, $k$ denotes the component of ${\bf{k}}$
along the chain direction, and the dynamical structure factor on a
chain with $L$ sites is defined as
\be
  S^{\alpha\alpha'}(\omega,k) =
   \frac{1}{L}\sum_{l,l'}
   \int\limits_{-\infty}^{\infty}\frac{dt}{2\pi}
   e^{i\omega t-ik(l-l')}
   \langle 0 |S_l^{\alpha}(t)S_{l'}^{\alpha'} | 0 \rangle.
  \label{s-alpha-alpha}
\ee
Substituting the low-energy expressions (\ref{CLS}) into
(\ref{s-alpha-alpha}) we obtain
\begin{eqnarray}
  S^{\alpha\alpha'}(\omega,k) &=&
  \sum_{a,b=1}^{3}\frac{1}{L}\sum_{l,l'}
   \int\limits_{-\infty}^{\infty}\frac{dt}{2\pi}
   e^{i\omega t-i(k-Q^\alpha_a)l+i(k+Q^{\alpha'}_{b})l'}\nn
&& \times\    \langle 0
|{\cal{S}}_{a}^{\alpha}(t,x){\cal{S}}_{b}^{\alpha'}(0,y)
   | 0 \rangle,
  \label{chi-alpha-alpha}
\end{eqnarray}
where $x=la_0$, $y=l'a_0$ and ${\cal{S}}_{a}^{\alpha}(x)$ are the
leading terms in the low energy limits (\ref{CLS}) of the lattice
spin operators. Using that the expectation value is a slowly
varying function of $x-y$ we see that only terms with \be k\approx
Q_a\approx -Q_b \ee contribute to
(\ref{chi-alpha-alpha})\cite{Wang}. The dynamical structure factor can
be expressed by means of a Lehmann representation in terms of
scattering states of solitons, anti-solitons and breathers. Inserting
a complete set of states (\ref{ROI}) between the operators in
(\ref{chi-alpha-alpha}) and using
\begin{eqnarray*}
  &&
  \langle 0 |
            {\cal{S}}_a^{\alpha}(t,x)
  |\{\epsilon_n,\theta_n\}\rangle=
  \\ && ~~~~~~~~~~ =
  e^{-iE_{\{n\}}t+iP_{\{n\}}x}
  \langle 0 |
            {\cal{S}}_a^{\alpha}(0,0)
  |\{\epsilon_n,\theta_n\}\rangle,
\end{eqnarray*}
we arrive at
\begin{widetext}
\be
  S^{\alpha\alpha'}(\omega,Q_a^{\alpha}+q)\simeq
  \frac{2\pi}{a_0}
   \sum_{n=1}^{\infty}
   \sum_{\{\epsilon_n\}}
   \sum_{b=1}^{3}\delta_{Q_a^\alpha,-Q_b^{\alpha'}}
   \int\frac{d\theta_1\ldots d\theta_n}
            {n! (2\pi)^n}
        \langle 0 |
              {\cal{S}}_a^{\alpha}
        | \{\epsilon_n,\theta_n\} \rangle
        \langle \{\epsilon_n,\theta_n\} |
              {\cal{S}}_{b}^{\alpha'}
        | 0 \rangle
        \delta(q-P_{\{n\}})\ \delta(\omega-E_{\{n\}})
   \label{chi-alpha-alpha-sp1}.
\ee
\end{widetext}
Here $q$ is assumed to be sufficiently small
($q\sim\frac{\Delta}{v}\ll{k}_F,\frac{\pi}{a_0}$). Due to
energy-momentum conservation only a finite number of intermediate
states contributes to the correlator
(\ref{chi-alpha-alpha-sp1}). Moreover, at low energies
contributions of intermediate states with large numbers of particles
to the correlator (\ref{chi-alpha-alpha-sp1}) are generally small
\cite{Cardy,CET}. We therefore restrict our following analysis to one
and two particle contributions. Many matrix elements in
(\ref{chi-alpha-alpha-sp1}) are in fact zero as can be established
by using charge conjugation symmetry and topological charge
conservation. The relevant properties of the continuum spin
operators ${\cal S}^\alpha_s$ are summarized in Table \ref{Ssymm}.
\begin{table}
\begin{tabular}{|c|c|c|c|c|c|c|c|c|c|}
  \hline
  & ${\cal S}^x_1$
  & ${\cal S}^x_2$
  & ${\cal S}^x_3$
  & ${\cal S}^y_1$
  & ${\cal S}^y_2$
  & ${\cal S}^y_3$
  & ${\cal S}^z_1$
  & ${\cal S}^z_2$
  & ${\cal S}^z_3$
  \\
  \hline \hline
  $Q$ & $-1$ & 0 & 1 & $-1$ & 0 & 1 & 0 & $-1$ &1
  \\
  \hline
  $C$ & & $+$ & &
  & $-$ & & $-$ & &
  \\
  \hline
\end{tabular}
\caption{Topological charge ${\cal{Q}}$ and eigenvalue (where
applicable) under charge conjugation $C$ of the continuum spin
operators.} \label{Ssymm}
\end{table}
Using these properties we furthermore conclude that at low
energies the non-vanishing components of the dynamical structure
factor are
\begin{enumerate}
\item $S^{xx}$, $S^{yy}$ in the vicinity of the
  points $k=\pm Q$;
\item $S^{xx}$, $S^{yy}$ near the point $k=\frac{\pi}{a_0}$;
\item $S^{zz}$ in the vicinity of the point $k=0$;
\item $S^{zz}$ near $k=\pm 2k_F$.
\end{enumerate}
In the following we determine these in the ``two-particle
approximation'', i.e., keeping only terms with $n\leq 2$ in the
spectral representation (\ref{chi-alpha-alpha-sp1}). In order to do so
we make use of the exact form of the matrix elements entering the
Lehmann representation, which follow from the form-factor
bootstrap approach \cite{FFBA,Babuji98}.

We note that as a consequence of charge conjugation symmetry the
components of the structure factor in the vicinities of $k=Q_a$
and $k=-Q_a$ are the same.

\section{Calculation of Correlation Functions: Kinematics}
\label{sec:kinematics}
The formalism we employ to calculate the dynamical structure factor
can be used quite generally to determine (real and imaginary parts of)
two-point correlation functions. The retarded two-point function of
two bosonic operators $A$ and $B$ has a spectral representations of
the form
\begin{widetext}
\begin{eqnarray}
  G^{AB}(\omega,q)&=&
  \frac{2\pi v}{a_0}
   \sum_{n=1}^{\infty}
   \sum_{\{\epsilon_n\}}
   \int\frac{d\theta_1\ldots d\theta_n}
            {n! (2\pi)^n}
   \bigg\{
        \langle 0 |
              A
        | \{\epsilon_n,\theta_n\} \rangle
        \langle \{\epsilon_n,\theta_n\} |
              B
        | 0 \rangle
        \frac{\delta(v q-P_{\{n\}})}
             {\omega-E_{\{n\}}+i\eta}    \nonumber    \\
&&\qquad\qquad\qquad -\
        \langle 0 |
              B
        | \{\epsilon_n,\theta_n\} \rangle
        \langle \{\epsilon_n,\theta_n\} |
              A
        | 0 \rangle
        \frac{\delta(vq+P_{\{n\}})}
             {\omega+E_{\{n\}}+i\eta}
   \bigg\}.
   \label{Gaa-sp1}
\end{eqnarray}
\end{widetext}
Here $\eta$ is a positive infinitesimal,
$|\{\epsilon_n,\theta_n\}\rangle$ are $n$-particle scattering
states of solitons, antisolitons and breathers
(\ref{n-particle-state}) with energies and momenta are given by
(\ref{P-E-n-particle}) and (\ref{P-E-epsilon}) respectively. The
leading contribution to the spectral sum in (\ref{Gaa-sp1}) is due
to intermediate states with one and two particles. Using momentum
conservation it is possible to simplify the expressions for these
contributions as we discuss next.

\subsection{One-particle kinematics}
Resolving the momentum conservation delta function leads to the
following result for the one-particle contributions to $G^{AB}$
\begin{widetext}
\begin{eqnarray}
  G^{AB}_{1\rm p}(\omega,q)&=&
  \frac{v}{a_0}
   \sum_{a}
   \int d\theta
   \bigg[ \langle 0 |A|\theta \rangle_a\
           {}_a\langle \theta | B | 0 \rangle
        \frac{\delta(v q-\Delta_a\sinh\th)}
             {\omega-\Delta_a\cosh\th+i\eta}
-\langle 0 | B | \theta\rangle_a\ {}_a\langle \theta | A | 0 \rangle
        \frac{\delta(vq+\Delta_a\sinh\th)}
             {\omega+\Delta_a\cosh\th+i\eta}   \bigg]\nonumber\\
&=&\sum_a\frac{v}{a_0\varepsilon_a(q)}\biggl[
\frac{\langle 0|A|\theta_0^a\rangle_a\ {}_a\langle \theta_0^a|B|0\rangle}
{\omega-\varepsilon_a(q)+i\eta}
-\frac{\langle 0|B|\theta_0^a\rangle_a\ {}_a\langle \theta_0^a|A|0\rangle}
{\omega+\varepsilon_a(q)+i\eta}
\biggr],
\end{eqnarray}
\end{widetext}
where $a$ runs over all single-particle labels (i.e. soliton,
antisoliton and breathers) and
\bea
\varepsilon_a(q)&=&\sqrt{\Delta_a^2+v^2q^2}\ ,\\
\theta^a_0&=&{\rm arcsinh}\Bigl(\frac{vq}{\Delta_a}\Bigr)\ .
\eea
\subsection{Two-particle kinematics}
As two-particle form factors of scalar operators depend only on
the rapidity difference, it is useful to change variables to
$\th_\pm=(\theta_1\pm\theta_2)/2$. Resolving the momentum
conservation delta function then gives
\begin{widetext}
\begin{eqnarray}
  G^{AB}_{2\rm p}(\omega,q)&=&
  \frac{v}{a_0}
  \sum_{a_1,a_2}
  \int \frac{d\theta_1 d\th_2}{2(2\pi)^2}
  \bigg[
       \langle 0 |
                 A
       |\theta_2,
        \theta_1 \rangle_{a_2a_1}
       {~}_{a_1a_2}\langle
        \theta_1,
        \theta_2 |
                 B
       | 0 \rangle
       \frac{\delta(v q-\sum_{j=1}^2\Delta_{a_j}\sinh\th_j)}
            {\omega-\sum_{j=1}^2\Delta_{a_j}\cosh\th_j+i\eta}
  \nonumber\\
  &&\qquad\qquad\qquad\qquad -
       \langle 0 |
                 B
       |\theta_2,
        \theta_1
       \rangle_{a_2a_1}
       {~}_{a_1a_2}\langle
        \theta_1,
        \theta_2 |
                A
       | 0 \rangle
       \frac{\delta(v q+\sum_{j=1}^2\Delta_{a_j}\sinh\th_j)}
            {\omega+\sum_{j=1}^2\Delta_{a_j}\cosh\th_j+i\eta}
  \bigg]
  \nonumber\\
  &=&
  \frac{v}{a_0}
  \sum_{a,b}
  \int\frac{d \theta_-}{2\pi}
  \biggl[
        \frac{\langle 0 |A|
              \th_0^{ab}-\th_-,\th_0^{ab}+\th_-\rangle_{ba}
              {~}_{ab}
              \langle\theta_0^{ab}+\th_-,\th_0^{ab}-\th_-
              |B | 0 \rangle}
             {\varepsilon_{ab}(q,\theta_-)\
              \big(\omega-\varepsilon_{ab}(q,\th_-)+
              i\eta\bigr)}
  \nonumber\\
  &&\qquad\qquad\qquad\qquad\qquad -
  \frac{\langle 0|B|\th_0^{ab}-\th_-,\th_0^{ab}+\th_- \rangle_{ba}\
  {}_{ab}\langle\theta_0^{ab}+\th_-,\th_0^{ab}-\th_- | A | 0\rangle}
  {\varepsilon_{ab}(q,\theta_-)\
  \bigl(
       \omega+\varepsilon_{ab}(q,\th_-)+i\eta\bigr)}
  \biggr],
  \label{G2p-def}
\end{eqnarray}
\end{widetext}
where
$$
 \varepsilon_{ab}(q,\theta)=
 \Bigl[
      v^2q^2+\Delta_a^2+\Delta_b^2+2\Delta_a\Delta_b\cosh(2\th)
 \Bigr]^\frac{1}{2},
$$
$$
 \theta_0^{ab}=
 \ln
 \left[
      \frac{vq+\varepsilon_{ab}(q,\th_-)}
      {\Delta_a\exp(\th_-)+\Delta_b\exp(-\th_-)}
 \right].
$$
The imaginary part of $G^{AB}_{2\rm p}(\omega,q)$ can be
simplified using
\bea
  -\frac{1}{\pi}
  {\rm Im}\
  \frac{1}{\varepsilon_{ab}(q,\th_-)\
  [\omega-\varepsilon_{ab}(q,\th_-)+i\eta]}
  ~~~~~~~~~~
  \nn
  =
  \frac{\delta(\th_--\frac{\th_{ab}(s)}{2})
  +\delta(\th_-+\frac{\th_{ab}(s)}{2})}
  {\sqrt{s^2-(\Delta_{a}+\Delta_{b})^2}\sqrt{s^2-(\Delta_{a}-\Delta_{b})^2}},
  \label{deltafns}
\eea
 where
\bea
  s^2 &=& \omega^2-v^2q^2\ ,
  \label{Mandelstams}\\
  \th_{ab}(s) &=&
  {\rm arccosh}
  \left[
       \frac{s^2-\Delta_a^2-\Delta_b^2}
            {2\Delta_a\Delta_b}
  \right] .
  \label{thab}
  \eea
 Carrying out the $\th_-$ integral using the delta functions we obtain
\begin{widetext}
\begin{eqnarray}
  -\frac{1}{\pi}{\rm Im}
  G^{AB}_{2\rm p}(\omega>0,q)
  &=&
  \frac{v}{2\pi a_0}
  \sum_{a,b}\sum_{\sigma=\pm}
  \frac{\langle0|
               A
        |\theta^{\sigma}_{ba}(\omega,q),
         \th^{-\sigma}_{ab}(\omega,q
        \rangle_{ba}
        {~}_{ab}
        \langle
         \th^{-\sigma}_{ab}(\omega,q),
         \theta^{\sigma}_{ba}(\omega,q)|
        B|0\rangle}
       {\sqrt{s^2-(\Delta_{a}+\Delta_{b})^2}
        \sqrt{s^2-(\Delta_{a}-\Delta_{b})^2}}
  \vartheta_H\bigl(s-\Delta_{a}-\Delta_{b}\bigr),
  \nn
  \label{imgaad}
\end{eqnarray}
\end{widetext}
where $\vartheta_H(x)$ is the Heaviside function and
\begin{eqnarray}
  &&
  \theta_{ab}^{\pm}(\omega,q)=
  {\rm arcsinh}
  \bigg[
       \frac{1}{2\Delta_{a}s^2}
       \Big(
           v q
           (s^2+\Delta_{a}^2-
            \Delta_{b}^2)+
  \nonumber \\ && ~~~ \pm\omega
          \sqrt{(s^2-(\Delta_{a}- \Delta_{b})^2)
                (s^2-(\Delta_{a}+ \Delta_{b})^2)}
       \Big)
  \bigg],
  \nonumber \\
  \label{theta12-omega-q}
\end{eqnarray}
$$
 \theta_{ab}^{\sigma}(\omega,q)-
 \theta_{ba}^{-\sigma}(\omega,q)=
 \sigma\theta_{ab}(s).
$$
The two terms in (\ref{imgaad}) arise from the two delta functions
in (\ref{deltafns}). Using the results summarized in this
section we can determine the one and two particle contributions to
both real and imaginary parts of two point functions. The
two particle contributions to the real part involve one (principal
part) integration, which is readily performed numerically.
In order to determine the dynamical structure factor we only require
the imaginary part of several two point functions.

\section{Results for the Dynamical Structure Factor}
        \label{sec:results}
Below we present results for the dynamical structure factor
$S^{\alpha\beta}(\omega,Q^{\alpha}_{a}+q)$
(\ref{chi-alpha-alpha-sp1})
in the regime $-1<\delta\leq 1$ and for magnetic fields
$H<H_c=J(1+\delta)$. We note that if $H\approx H_c$ or
$\delta\approx -1$ the cut-off in the field theory is very
small, which limits the utility of our approach. For the sake of
clarity we use a particular set of parameters in all plots
\begin{equation}
  \gamma=\frac{h}{H}=0.01191,
  \ \ \
  \delta=0.3,
  \ \ \
  H=0.2598 J.
\end{equation}
These correspond to a magnetization per site of
$\langle{S^z}\rangle=0.06$ (see Table \ref{table-m-a-c-b-beta-H})
and $\xi=0.174371$. The spectrum consists of soliton and
antisoliton with gap $\Delta\approx0.04897J$ and five breathers
with gaps
\begin{eqnarray}
\Delta_1&=&0.54098\Delta\ ,\ \Delta_2=1.04162\Delta\ ,\
\Delta_3=1.46461\Delta,\nonumber\\
\Delta_4&=&1.77841\Delta\ ,\ \Delta_5=1.95962\Delta.
\end{eqnarray}
In order to broaden delta functions appearing in one particle
contributions, we introduce a small imaginary part in $\omega$,
equal to $\eta=0.01\Delta$.

\subsection{$S^{xx}(\omega,k)$}
    \label{Subsec-xx}
In the continuum limit $S^{xx}(\omega,k)$ is non-vanishing in the
vicinity of the points $k=\pm Q$ and $\frac{\pi}{a_0}$. We will
consider both cases in turn. As we have noted before, the response at
$k=\pm Q$ is indentical as a result of charge conjugation symmetry, so
that it is sufficient to consider $k\approx-Q$.

\subsubsection{Momenta $k\approx-Q=-\frac{2\pi}{a_0}\langle S^z_j\rangle$}

In the continuum limit $S^{xx}(\omega,-Q+q)$ with $q\ll Q$ is given
by the two-point function of ${\cal S}_3^x$ with ${\cal S}_1^x$
(\ref{Sx-smooth-scalar}).
This is because $vQ$ is a large energy scale proportional to the
cutoff in the theory. Using Table \ref{Ssymm} we find that the
following intermediate states with at most two particles
contribute
\begin{enumerate}
\item{} Single-soliton states.
\item{} Two particle states containing one soliton and one breather.
\end{enumerate}
The corresponding matrix elements are calculated in Appendix
\ref{Sec-exp-FF-sb}. Using the results of section \ref{sec:kinematics}
to carry out the rapidity integrals we arrive at the following
expression for $S^{xx}(\omega,-Q+q)$ within the two-particle
approximation
\begin{widetext}
\begin{eqnarray}
  S^{xx}(\omega>0,-Q+q)&\approx&
  \frac{2v\tilde{{\cal A}}^2}{a_0}
  \frac{v^2q^2}{\Delta^2}
  \delta(s^2-\Delta^2)+
   \frac{v\tilde{{\cal A}}^2}{4\pi a_0}
  \sum_{k=1}^{[1/\xi]}
  \left(N^{\beta}_{sb_k}\right)^2
  \frac{\left|
             F^{\rm{min}}_{sb_k}(\theta_{sb_k}(s))
        \right|^2
        \vartheta_H(s-\Delta-\Delta_k)}
       {\sqrt{(s^2-(\Delta-\Delta_k)^2)
              (s^2-(\Delta+\Delta_k)^2)}}
  \times \nonumber \\ && ~~~~~~~~ \times
  \sum_{\sigma=\pm}
  \Big[
      K^{\beta}_{sb_k}(\sigma\theta_{sb_k}(s))
      e^{\theta_{sb_k}^{\sigma}(\omega,q)}+
      K^{-\beta}_{sb_k}(\sigma\theta_{sb_k}(s))
      e^{-\theta_{sb_k}^{\sigma}(\omega,q)}
  \Big]^2.
  \label{G-xx}
\end{eqnarray}
\end{widetext}
Here $s$ is the Mandelstam variable (\ref{Mandelstams}), the
overall normalization is
\be
 \tilde{{\cal A}}=
 {\cal A}(H)
 \bigg(
      Z_1(\beta)
 \bigg)^{1/2},
 \label{tilde-A}
\ee where $Z_1(\beta)$ is given by equation (\ref{Zn}), the
minimal form factors $F^{\rm{min}}_{sb_k}(\theta)$ by equation
(\ref{F-sbn-min}), the pole functions $K^{\pm}_{sb_k}(\theta)$ by
equation (\ref{K-2m}) for $k$ even and (\ref{K-2m+1}) for $k$ odd,
the normalization factor $N^{\beta}_{sb_k}$ by equation
(\ref{N-sbn}) and the functions $\theta_{sb_k}(s)$ and
$\theta^{\sigma}_{sb_k}(\omega,q)$ are presented in equations
(\ref{thab}) and (\ref{theta12-omega-q}), respectively.

\begin{figure}[htb]
\centering
\includegraphics[width=60mm,height=40mm,angle=0]{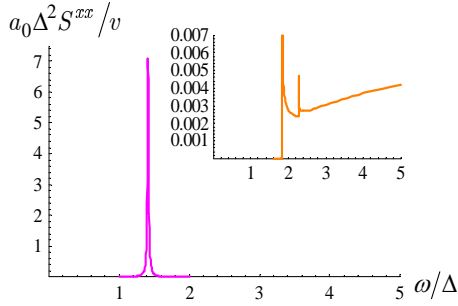}
 \caption{One and two-particle contributions to
   $S^{xx}(\omega,-Q+\Delta/v)$ as a function of $\omega$
   for $\delta=0.3$ and $H=0.2598 J$. The delta-function peak
   (pink) has been broadened to make it visible.}
 \label{Fig-chi-xx-Q}
\end{figure}
We note that $S^{xx}(\omega,-Q+q)$ vanishes when $q\to0$. In
Figure \ref{Fig-chi-xx-Q} we therefore plot $S^{xx}(\omega,
-Q+\Delta/v)$ as a function of $\omega$. In order to broaden
delta-function contributions we introduce a small imaginary part
in $\omega$. Two features are clearly visible: there is a coherent
peak corresponding to the contribution of single-soliton
excitations at energy at $\Delta\sqrt{2}$. At higher energies
breather-soliton continua appear. Their contributions grow
with increasing $\omega$ because ${\cal S}_3^x$ is an irrelevant
operator. It is instructive to compare our result to the gapless
spin-1/2 Heisenberg XXZ chain, see e.g. Ref.[\onlinecite{citra}].
There one has
 \be
  S^{xx}(\omega>0,-Q+q) \propto
  \frac{\omega^2+v^2q^2}{(\omega^2-v^2q^2)^{1-\nu}},
 \ee
where $\nu=2(\beta+\frac{1}{4\beta})^2>1$. For large $\omega$ this
increases as $\omega^{2\nu}$, while is goes to zero in a power-law
fashion for $\omega\to vq$. In presence of a staggered field, the
dynamical structure factor (\ref{G-xx}) has divergence for
$\omega\to \sqrt{(\Delta+\Delta_k)^2+v^2q^2}$
($k=1,2,\ldots,[1/\xi]$), while the large frequency behavior is
the same as without the staggered field.
\subsubsection{Vicinity of antiferromagnetic wave number:
 $k\approx \pi/a_0$}
In the continuum limit $S^{xx}(\omega,\frac{\pi}{a_0}+q)$ with
$qa_0\ll\pi$ is given by the two-point function of the charge
neutral operator ${\cal S}^x_2$ (\ref{Sx-stag-scalar}). Using
Table \ref{Ssymm} and (\ref{charge-conjugate}) we find that the
following intermediate states with at most two particles
contribute to the two-point function of ${\cal S}^x_2$
\begin{enumerate}
\item{} Single breather states even under charge conjugation,
  i.e. $B^\dagger_{2n}(\theta)|0\rangle$.
\item{} Two particle states containing one soliton and one
  antisoliton.
\item{} Two particle states containing two even or two odd breathers.
\end{enumerate}
Using the results of section \ref{sec:kinematics}, we obtain the
following expression in the two-particle approximation
\begin{widetext}
\begin{eqnarray}
  S^{xx}\left(\omega>0,\frac{\pi}{a_0}+q\right) &\approx&\frac{vc^2(H)}{\pi a_0}\Biggl\{
  2\pi
  \sum_{k=1}^{[1/2\xi]}
  \left|F^{\beta}_{b_{2k}}\right|^2
  \delta(s^2-\Delta_{2k}^2)+
  \frac{\big|
            F^{\cos(\beta\Theta)}_{s\bar s}(\theta_{s\bar s}(s))
        \big|^2
        \vartheta_H(s-2\Delta)}
       {s\sqrt{s^2-4\Delta^2}}+
  \nonumber \\ && +
  \sum_{k,k'=1}^{[1/\xi]}
  \delta_{k+k'}^{\rm{even}}
  \big|
      F^{\beta}_{b_kb_{k'}}(\theta_{b_kb_{k'}}(s))
  \big|^2
  \frac{\vartheta_H(s-\Delta_k-\Delta_{k'})}
       {\sqrt{(s^2-(\Delta_k-\Delta_{k'})^2)
              (s^2-(\Delta_k+\Delta_{k'})^2)}}\Biggr\}.
  \label{Dxx}
\end{eqnarray}
\end{widetext}
Here the single-breather form factors $F^{\beta}_{b_k}$ are given
by equation (\ref{FF-bn-beta}), the soliton-antisoliton form
factor $F_{s\bar{s}}^{\cos(\beta\Theta)}(\theta)$ by
(\ref{FF-sa-cos}) and the breather-breather form factors
$F^{\beta}_{b_kb_{k'}}(\theta)$ by (\ref{FF-beta-kl})
respectively. The function $\theta_{\epsilon\epsilon'}(s)$ is
given by (\ref{thab}) and
\begin{equation}
  \delta_{k}^{\rm{even}}=
  \left\{
  \begin{array}{ll}
  1, & {\rm{if~}} k {\rm{~is~even}},
  \\
  0, & {\rm{overwise}}.
  \end{array}
  \right.
  \label{delta-k-even}
\end{equation}

\begin{figure}[htb]
\centering
\includegraphics[width=60mm,height=39mm,angle=0]{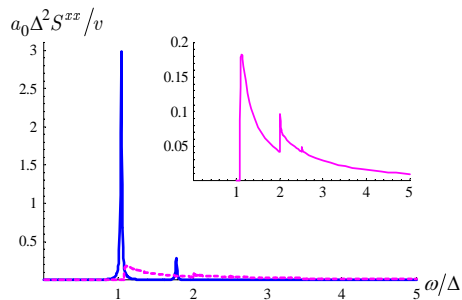}
 \caption{One and two-particle contributions to
 $S^{xx}(\omega,\frac{\pi}{a_0})$ as a function of $\omega$
 for $\delta=0.3$ and $H=0.2598 J$. Delta-function peaks (blue)
 have been broadened to make them visible. }
 \label{Fig-chi-xx-pi}
\end{figure}
In Fig.\ref{Fig-chi-xx-pi} we plot the dynamical structure factor
(\ref{Dxx}) as a function of frequency. We note that because
${\cal S}_2^x$ is a scalar operator
$S^{xx}(\omega,\frac{\pi}{a_0}+q)$ depends only on the
Mandelstam variable $s$ (\ref{Mandelstams}) rather than on
$\omega$ and $q$ separately. The first peak in
$S^{xx}(\omega,\frac{\pi}{a_0})$ is due to the $b_2$
single-breather excitation (blue line). At $\omega=\Delta_4$ there
is a second single-breather contribution, due to $b_4$. Above
$\omega=2\Delta_1$ a strong $b_1b_1$ two-breather continuum occurs
(pink line). Around $\omega=2\Delta$ contributions from
soliton-antisoliton and $b_1b_3$ and $b_2b_2$ two-breather
continua are visible. We note that the thresholds of $b_1b_3$,
$s\bar{s}$ and $b_2b_2$ continua all occur around $2\Delta$ is a
peculiarity of the parameters we have chosen in the plots.

\subsection{$S^{yy}(\omega,k)$}
    \label{Subsec-yy}
Next we turn to the $yy$-component of the dynamical structure
factor. In the continuum limit $S^{yy}(\omega,k)$ is non-vanishing
in the vicinity of the points $k=\pm Q$ and $\frac{\pi}{a_0}$. We
will consider both cases in turn.
\subsubsection{Momenta $k\approx-Q=-\frac{2\pi}{a_0}\langle S^z_j\rangle$}
In the continuum limit $S^{yy}(\omega,-Q+q)$ with $q\ll Q$ is given
by the two-point function of ${\cal S}^y_3$ with ${\cal S}^y_1$
(\ref{Sy-smooth-scalar}). Using Table \ref{Ssymm} we find that the
following intermediate states with at most two particles
contribute to the two-point function
\begin{enumerate}
\item{} Single-soliton states.
\item{} Two particle states containing one soliton and one breather.
\end{enumerate}
The corresponding matrix elements are calculated in Appendix
\ref{Sec-exp-FF-sb}. Carrying out the rapidity integrals, see
section \ref{sec:kinematics}, we arrive at the following
expression for $S^{yy}(\omega,-Q+q)$ within the two-particle
approximation
\begin{widetext}
\begin{eqnarray}
  S^{yy}(\omega>0,-Q+q) &\approx&
  \frac{2v\tilde{{\cal A}}^2}{a_0}
  \frac{v^2q^2+\Delta^2}{\Delta^2}
  \delta(s^2-\Delta^2)+
  \frac{v\widetilde{\cal{A}}^2}{4\pi a_0}
  \sum_{k=1}^{[1/\xi]}
  \Big(N^{\beta}_{sb_k}\Big)^2
  \frac{\left|
             F^{\rm min}_{sb_k}(\theta_{sb_k}(s))
        \right|^2
        \vartheta_H(s-\Delta-\Delta_k)}
       {\sqrt{(s^2-(\Delta-\Delta_k)^2)
              (s^2-(\Delta+\Delta_k)^2)}}
  \times \nonumber \\ && \times
  \sum_{\sigma=\pm}
  \Big[
      K^{\beta}_{sb_k}(\sigma\theta_{sb_k}(s))
      e^{\theta_{sb_k}^{\sigma}(\omega,q)}-
      K^{\beta}_{sb_k}(-\sigma\theta_{sb_k}(s))
      e^{-\theta_{sb_k}^{\sigma}(\omega,q)}
  \Big]^2.
  \label{G-yy}
\end{eqnarray}
\end{widetext}
Here the overall normalization $\widetilde{\cal{A}}$ is given by
equation (\ref{tilde-A}), the minimal form factor
$F^{\rm{min}}_{sb_k}(\theta)$ by (\ref{F-sbn-min}), the pole
function $K^{\beta}_{sb_k}(\theta)$ by (\ref{K-2m}) for $k$ even
and (\ref{K-2m+1}) for $k$ odd, the functions
$\theta_{sb_k}(s)$ and $\theta^{\sigma}_{sb_k}(\omega,q)$ by
(\ref{thab}) and (\ref{theta12-omega-q}), respectively.

\begin{figure}[htb]
\centering
\includegraphics[width=60mm,height=40mm,angle=0]{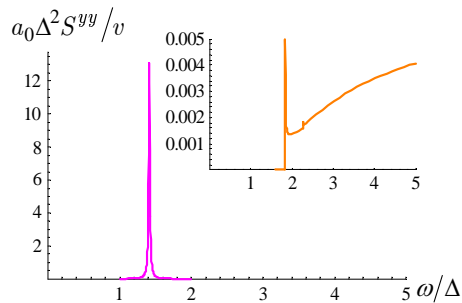}
 \caption{One and two-particle contributions to
   $S^{yy}(\omega,-Q+\Delta/v)$ as a function of $\omega$
   for $\delta=0.3$ and $H=0.2598 J$. The delta-function peak (pink)
   has been broadened to make it visible. Inset: the
   soliton-breather two-particle contributions.}
 \label{Fig-chi-yy-Q}
\end{figure}
We plot
$S^{yy}(\omega,-Q+\Delta/v)$ as a function of $\omega$ in Figure
\ref{Fig-chi-yy-Q}. Delta-function contributions have been
broadened to make them visible. We see that there is a coherent
peak corresponding to the contribution of single-soliton
excitations at energy at $\Delta\sqrt{2}$. At higher energies
breather-soliton continua appear. Their contributions grow
with increasing $\omega$ because ${\cal S}_3^y$ is an irrelevant
operator.

\subsubsection{Vicinity of antiferromagnetic wave number:
 $k\approx \pi/a_0$}
In the continuum limit $S^{yy}(\omega,\frac{\pi}{a_0}+q)$ with
$qa_0\ll\pi$ is given by the two-point function of the charge
neutral operator ${\cal S}^y_2$ (\ref{Sy-stag-scalar}). Using
Table \ref{Ssymm} and (\ref{charge-conjugate}) we find that the
following intermediate states with at most two particles
contribute to the two-point function of ${\cal S}^y_2$
\begin{enumerate}
\item{} Single breather states odd under charge conjugation,
  i.e. $B^\dagger_{2n+1}(\theta)|0\rangle$.
\item{} Two particle states containing one soliton and one
  antisoliton.
\item{} Two particle states containing one even and one odd breather.
\end{enumerate}
Using the results of section \ref{sec:kinematics}, we obtain the
following expression in the two-particle approximation
\begin{widetext}
\begin{eqnarray}
  S^{yy}\left(\omega>0,\frac{\pi}{a_0}+q\right) &\approx&\frac{vc^2(H)}{\pi a_0}\Biggl\{
  2\pi
  \sum_{k=1}^{[1/\xi]}
  \delta_{k}^{\rm{odd}}
  \left|F^{\beta}_{b_k}\right|^2
  \delta(s^2-\Delta_k^2)+
  \frac{\big|
            F^{\sin(\beta\Theta)}_{s\bar s}(\theta_{s\bar s}(s))
        \big|^2
        \vartheta_H(s-2\Delta)}
       {s\sqrt{s^2-4\Delta^2}}+
  \nonumber \\ && +
  \sum_{k,k'=1}^{[1/\xi]}
  \delta_{k+k'}^{\rm{odd}}
  \big|
      F^{\beta}_{b_kb_{k'}}(\theta_{b_kb_{k'}}(s))
  \big|^2
  \frac{\vartheta_H(s-\Delta_k-\Delta_{k'})}
       {\sqrt{(s^2-(\Delta_k-\Delta_{k'})^2)
              (s^2-(\Delta_k+\Delta_{k'})^2)}}\Biggr\}.
  \label{D-yy}
\end{eqnarray}
\end{widetext}
Here the single-breather form factors $F^{\beta}_{b_k}$ are given
by (\ref{FF-bn-beta}), the soliton antisoliton form factor
$F_{s\bar{s}}^{\sin(\beta\Theta)}(\theta)$ by (\ref{FF-sa-sin}),
the two-breather form factors $F^{\beta}_{b_kb_{k'}}(\theta)$ by
(\ref{FF-beta-kl}), the function $\theta_{ab}(s)$ by (\ref{thab})
and
\begin{equation}
  \delta_{k}^{\rm{odd}}=
  \left\{
  \begin{array}{ll}
  1, & {\rm{if~}} k {\rm{~is~odd}},
  \\
  0, & {\rm{overwise}}.
  \end{array}
  \right.
  \label{delta-k-odd}
\end{equation}

\begin{figure}[htb]
\centering
\includegraphics[width=60mm,height=38mm,angle=0]{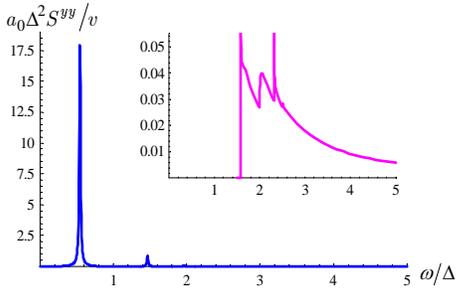}
 \caption{One and two-particle contributions to
 $S^{yy}(\omega,\frac{\pi}{a_0})$ as a function of $\omega$ for
 $\delta=0.3$ and $H=0.2598 J$. Delta-function peaks (blue) have
 been broadened to make them visible. Insert: the
 soliton-antisoliton and breather-breather two-particle
 contributions.}
 \label{Fig-chi-yy-pi}
\end{figure}
We plot $S^{yy}(\omega,\frac{\pi}{a_0})$ as a function of $\omega$
in Fig.\ref{Fig-chi-yy-pi}. We see that it is dominated
by the contribution of the first breather $b_1$ (the corresponding
delta function has been broadened). The contributions from $b_3$
and $b_5$ single-breather states are small in comparison.
Similarly, the two-particle $b_1b_2$, $s{\bar{s}}$ and $b_1b_4$
continua shown in the inset of Fig.\ref{Fig-chi-yy-pi} are
negligible.
\subsection{Longitudinal structure factor $S^{zz}(\omega,k)$}
    \label{Subsec-zz}
We now consider the $zz$-component of dynamical structure factor.
In the continual limit $S^{zz}(\omega,k)$ is non-vanishing in the
vicinity of the points $k=0$ and $\pm{2k_F}$. We will consider
both cases in turn.

\subsubsection{Vicinity of ferromagnetic wave number:
 $k\approx 0$}
In the continuum limit $S^{zz}(\omega,q)$ with $qa_0\ll\pi$ is
given by the two-point function of the charge neutral operator
${\cal S}^z_1$ (\ref{Sz-smooth-scalar}). Using Table \ref{Ssymm}
and (\ref{charge-conjugate}) we find that the following
intermediate states with at most two particles contribute to the
two-point function of ${\cal S}^z_1$
\begin{enumerate}
\item{} Single breather states odd under charge conjugation,
  i.e. $B^\dagger_{2n+1}(\theta)|0\rangle$.
\item{} Two particle states containing one soliton and one
  antisoliton.
\item{} Two particle states containing one even and one odd breather.
\end{enumerate}
Using the results of section \ref{sec:kinematics}, we obtain the
following expression in the two-particle approximation
\begin{widetext}
\begin{eqnarray}
  S^{zz}(\omega>0,q) &\approx&
  \frac{2a_0\widetilde{b}^2\omega^2}{v}
  \sum_{k=1}^{[1/\xi]}
  \delta_{k}^{\rm{odd}}
  \Big|
      F_{b_k}^{\Theta}
  \Big|^2
  \delta(s^2-\Delta_k^2)+
  \frac{a_0\widetilde{b}^2\omega^2}{v}
  \frac{\big|
            F^{\Theta}_{s\bar s}(\theta_{s\bar s}(s))
        \big|^2
        \vartheta_H(s-2\Delta)}
        {s\sqrt{s^2-4\Delta^2}}+
  \nonumber \\ && +
  \frac{a_0\widetilde{b}^2\omega^2}{v}
  \sum_{k,k'=1}^{[1/\xi]}
  \delta_{k+k'}^{\rm{odd}}
  \big|
      F^{\Theta}_{b_kb_{k'}}(\theta_{b_kb_{k'}}(s))
  \big|^2
  \frac{\vartheta_H(s-\Delta_k-\Delta_{k'})}
       {\sqrt{(s^2-(\Delta_k-\Delta_{k'})^2)
              (s^2-(\Delta_k+\Delta_{k'})^2)}},
  \label{S-0-zz}
\end{eqnarray}
\end{widetext}
where the single-breather form factor $F^{\Theta}_{b_k}$ is given
by equation (\ref{FF-Phi-bn}), the soliton-antisoliton form factor
$F_{s\bar{s}}^{\Theta}(\theta)$ by equation (\ref{FF-sa-Phi}), the
breather-breather form factor $F^{\Theta}_{b_kb_{k'}}(\theta)$ by
equation (\ref{FF-Phi-kl}), $\theta_{ab}(s)$ is given by equation
(\ref{thab}), $s$ is the Mandelstam variable (\ref{Mandelstams}),
$\delta_k^{\rm{odd}}$ is given in (\ref{delta-k-odd})
and the overall normalization is
\be
 \widetilde{b}=
 \frac{1}{4\pi\beta}.
 \label{tilde-b}
\ee

\begin{figure}[htb]
\centering
\includegraphics[width=60mm,height=41mm,angle=0]{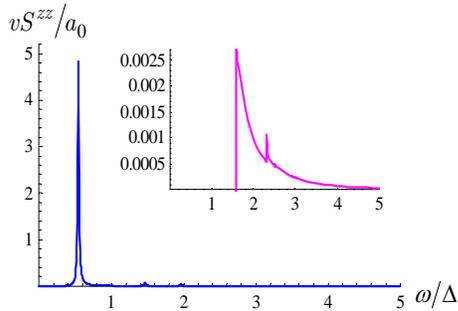}
 \caption{One and two particle contribution to $S^{zz}(\omega,0)$
 (\ref{S-0-zz}) as a function of $\omega$ for $\delta=0.3$ and
 $H=0.2598 J$. Delta-function peaks (blue) have been broadened
 to make them visible. Insert: the soliton-antisoliton and
 breather-breather two-particle contributions.}
 \label{Fig-chi-zz-0}
\end{figure}

The dynamical structure factor (\ref{S-0-zz}) is shown in Figure
\ref{Fig-chi-zz-0}. Note that since ${\cal{S}}_1^z$ is a scalar
operator, $S^{zz}(\omega,q)$ depends on the Mandelstam variable
$s$ (\ref{Mandelstams}) rather than on $\omega$ and $q$
separately. In order to broaden the delta function contributions
we introduce a small imaginary part in $\omega$. The dominant peak in
$S^{zz}(\omega,q)$ is due to a $b_1$ breather contribution. The
contributions due to $b_3$ and $b_5$  breather states are much
smaller. The  soliton-antisoliton and breather-breather
contributions to $S^{zz}(\omega,q)$ are barely visible in the figure.

\subsubsection{Momenta $k\approx-2k_F$}
In the continuum limit $S^{zz}(\omega,-2k_F+q)$ with $qa_0\ll\pi$
is given by the two-point function of ${\cal S}^z_3$ with ${\cal S}^z_2$
(\ref{Sz-stag-scalar}). Using Table \ref{Ssymm} and
(\ref{charge-conjugate}) we find that the following intermediate
states with at most two particles contribute to the two-point
function of
\begin{enumerate}
\item{} Single soliton state.
\item{} Two particle states containing one soliton and one
  breather.
\end{enumerate}
Using the results of section \ref{sec:kinematics}, we obtain the
following expression in the two-particle approximation
\begin{widetext}
\begin{eqnarray}
  S^{zz}(\omega>0,-2k_F+q)\approx
  \frac{v\tilde{a}^2}{a_0}
  \delta(s^2-\Delta^2)+
  \frac{v\tilde{a}^2}{2\pi a_0}
  \sum_{k=1}^{[1/\xi]}
  \Big(N^0_{sb_k}\Big)^2
  \Big(
      K^0_{sb_k}(\theta_{sb_k}(s))
  \Big)^2
  \frac{\Big|
            F^{\rm min}_{sb_k}(\theta_{sb_k}(s))
        \Big|^2
        \vartheta_H(s-\Delta-\Delta_k)}
       {\sqrt{(s^2-(\Delta-\Delta_k)^2)
              (s^2-(\Delta+\Delta_k)^2)}}.
  &&
  \nonumber \\
  \label{G-zz}
\end{eqnarray}
\end{widetext}
Here the minimal form factor $F^{\rm{min}}_{sb_k}(\theta)$ is
given by (\ref{F-sbn-min}), the pole function
$K^{0}_{sb_k}(\theta)$ by (\ref{K-2m}) for $k$ even and
(\ref{K-2m+1}) for $k$ odd, the function $\theta_{sb_k}(s)$ by
(\ref{thab}), the overall normalization is \be
  \tilde{a}=
  a(H)
  \sqrt{\frac{Z(0)}{2}}.
  \label{tilde-little-a}
\ee

\begin{figure}[htb]
\centering
\includegraphics[width=60mm,height=39mm,angle=0]{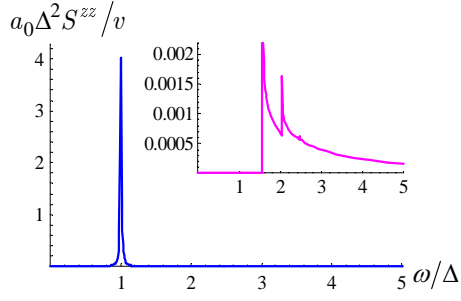}
 \caption{One and two particle contribution to $S^{zz}(\omega,-2k_F)$
 (\ref{G-zz}) as a function of $\omega$ for $\delta=0.3$ and
 $H=0.2598 J$. Delta-function peak (blue) has been broadened
 to make it visible. Insert: the soliton-breather two particle
 contributions.}
 \label{Fig-chi-zz-2kF}
\end{figure}

The dynamical structure factor (\ref{G-zz}) is shown in Figure
\ref{Fig-chi-zz-2kF}. Here we chose $q=0$. The strong low-energy peak
in $S^{zz}(\omega,-2k_F)$ is due to a one-soliton
state. Soliton-breather continua appear at higher energies.

\section{Summary and Conclusions}
        \label{sec:conc}
In this work we have determined the low energy dynamical spin response
of the anisotropic spin-1/2 Heisenberg XXZ chain in the presence of
both uniform and stagered magnetic fields. The uniform field was taken
to be along the anisotropy axis and the staggered field perpendicular
to it. The qualitative features of the model such as a field induced
gap and the formation of bound states are similar to the case of
isotropic exchange, which has been previously studied in detail
\cite{oa97,ET97,Essler99,oa99,oa02,Lou02,Wolter03b,zhao03,lou05}. The
main effect of a strong exchange anisotropy is to generate further
bound states and increase the binding energy. We have analyzed these
effects on the dynamic response and determined for the first time all
two-particle contributions, in particular those containing one soliton
and one breather. The results obtained here can be used to
study a quasi one dimensional array of anisotropic Heisenberg chains
in a uniform magnetic field by combining a mean-field approach with an
RPA-like approximation \cite{Schulz,EsslerCo-97,andrey}. This is of interest
in view of neutron scattering experiments on the quasi-1D anisotropic
Heisenberg magnet ${\rm Cs_2CoCl_4}$ \cite{KenzelmannPRB65}.

\section{ACKNOWLEDGMENT}
        \label{sec:akn}

This work was supported by the EPSRC under Grant No.
EP/D0500952/1.

\appendix

\section{Spin Velocity, Fermi Momentum and Compactification Radius}
\label{app:BA}
\begin{figure}[ht]
\begin{center}
\epsfxsize=0.45\textwidth \epsfbox{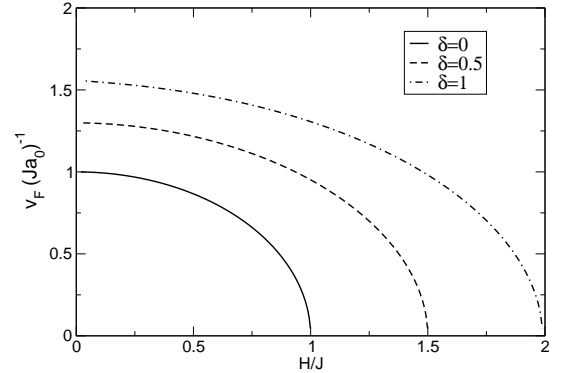}\quad
\end{center}
\caption{Spin velocity as a function of magnetic field for
different
  values of $\delta$}
\label{fig:vf}
\end{figure}

\begin{figure}[ht]
\begin{center}
\epsfxsize=0.45\textwidth \epsfbox{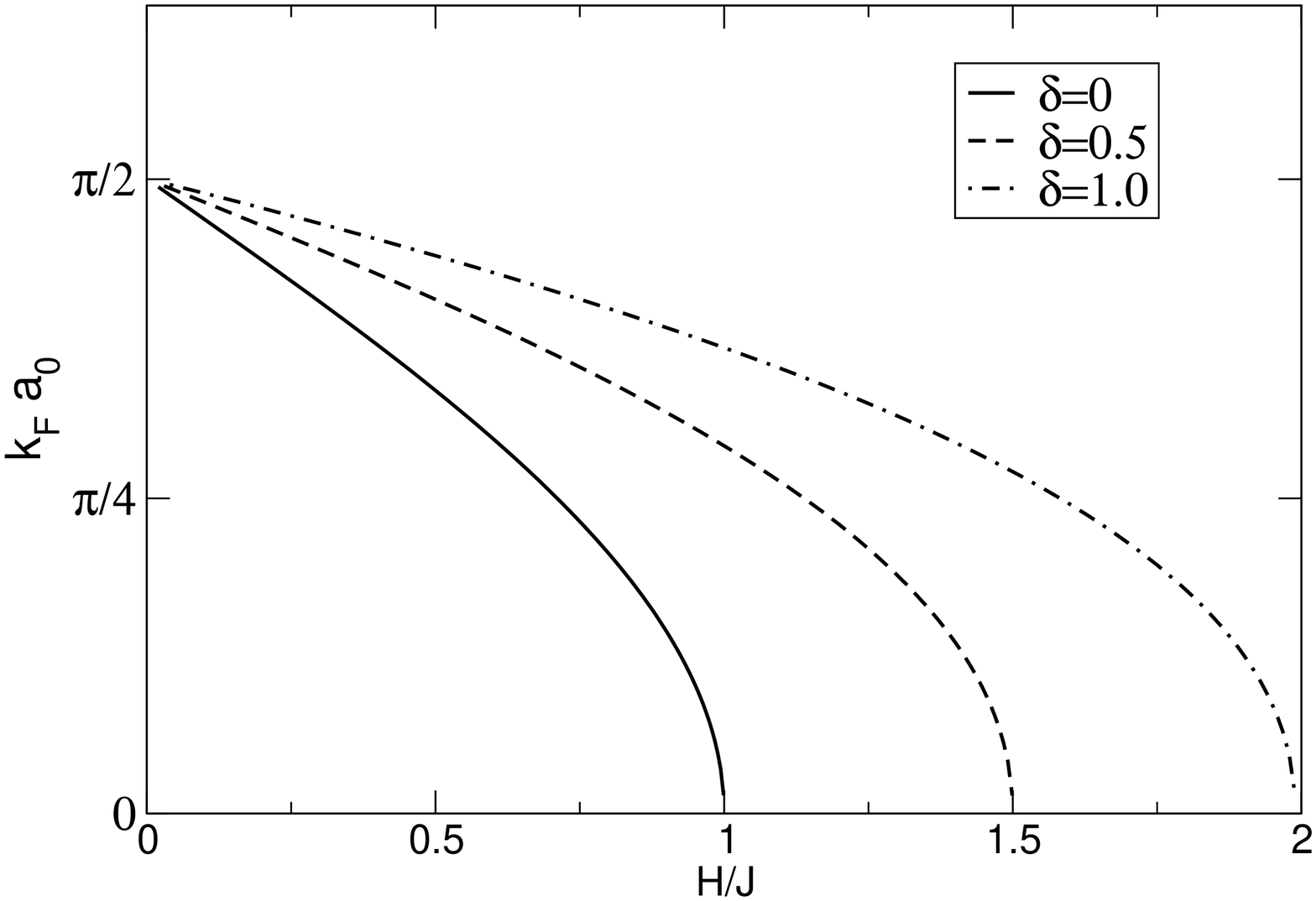}\quad
\end{center}
\caption{``Fermi momentum'' $k_F$  as a function of magnetic field
for different values of $\delta$} \label{fig:kf}
\end{figure}

\begin{figure}[ht]
\begin{center}
\epsfxsize=0.45\textwidth \epsfbox{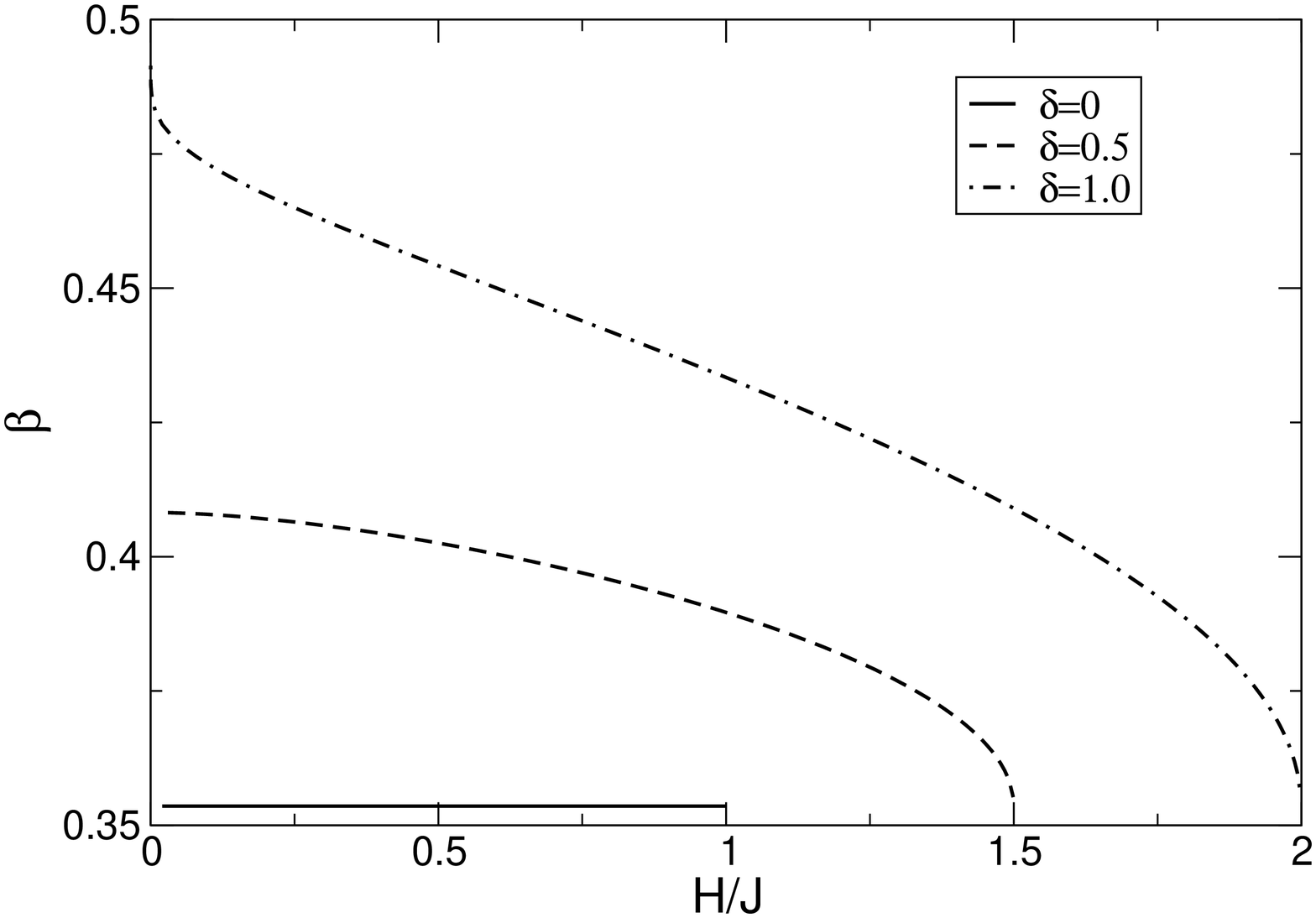}\quad
\end{center}
\caption{Parameter $\beta$ as a function of magnetic field for
different
  values of $\delta$}
\label{fig:beta}
\end{figure}

In this appendix we summarize how to determine the parameters of
the Gaussian model (\ref{GaussianModel}), (\ref{compactTheta})
that describes the continuum limit of the Heisenberg XXZ chain in
a magnetic field from the Bethe ansatz solution \cite{vladb}. The
velocity, Fermi momentum and compatification radius are expressed
in terms of the solutions of the following set of linear integral
equations for the dressed energy $\eps(\lambda)$, dressed momentum
$p(\lambda)$, dressed density $\rho(\lambda)$ and dressed charge
$Z(\lambda)$
\bea \eps(\lambda)&-&\!\int_{-A}^A \frac{d\mu}{2\pi}\
K(\lambda-\mu)\ \eps(\mu) = H-\frac{J\sin^2\gamma}{\cosh 2\lambda
-\cos\gamma},\nn p(\lambda)&=&\frac{2\pi}{a_0}\int_0^\lambda d\mu\
\rho(\mu)\ ,\nn \rho(\lambda)&-&\int_{-A}^A \frac{d\mu}{2\pi}\
K(\lambda-\mu)\ \rho(\mu) = \frac{2\sin\gamma}{2\pi[\cosh 2\lambda
-\cos\gamma]}\ ,\nn Z(\lambda)&-&\int_{-A}^A \frac{d\mu}{2\pi}\
K(\lambda-\mu)\ Z(\mu) = 1\ . \label{inteqs}
\eea
Here the exchange anisotropy is parametrized as $\delta=\cos(\gamma)$
and the integral kernel is given by
\be K(\lambda)=-2\sin 2\gamma/(\cosh2\lambda -\cos
2\gamma).
\ee
The integration boundary $A$ is fixed by the
condition
\be
\eps(\pm A)=0\ .
\ee
The physical meaning of the
various quantities is as follows: $\eps(\lambda)$ and $p(\lambda)$
are the energy and momentum of an elementary ``spinon'' excitation
carrying spin $S^z=\pm\frac{1}{2}$. We note that spinons can only
be excited in pairs. The magnetization per site in the ground
state is given in terms of the ground state root density
$\rho(\lambda)$ as \be \langle
S^z_j\rangle=\frac{1}{2}-\int_{-A}^Ad\lambda\ \rho(\lambda) \ee
The Fermi momentum is equal to \be
k_F=p(A)=\frac{2\pi}{a_0}\int_0^A d\lambda\ \rho(\lambda)
=\frac{\pi}{a_0}\left[\frac{1}{2}-\langle S^z_j\rangle\right], \ee
where we have used that $\rho(-\lambda)=\rho(\lambda)$. The spin
velocity is equal to the derivative of the spinon energy with
respect to the momentum at the Fermi points \be
v=\frac{\partial\epsilon(\lambda)}{\partial
p(\lambda)}\bigg|_{\lambda=A}=\frac{\partial\epsilon(\lambda)/\partial\lambda}
{2\pi\rho(\lambda)}\bigg|_{\lambda=A} a_0\ . \label{velocity} \ee
Finally, the dressed charge is related to $\beta$ by \be \beta=
\frac{1}{\sqrt{8}Z(A)}\ . \label{beta} \ee In order to determine
$v$ and $\beta$ we solve (\ref{inteqs}) numerically, which is
easily done to very high precision as the equations are linear.
The results are shown in Fig.\,\ref{fig:vf}, \ref{fig:kf} and
\ref{fig:beta}.

For zero magnetic field we have $k_F=\pi/2a_0$,
$\beta^2=\frac{1}{4\pi}\,{\rm arccos}(-\delta)$ and
\bea
v=\frac{Ja_0}{2}\,\frac{\sin4\pi\beta^2}{1-4\beta^2}\,.
\eea

\section{Form factors of the exponential field ${\cal{O}}^{-1}_a$}
        \label{Sec-exp-FF-sb}
The exponential field ${\cal{O}}^{-1}_a$ (\ref{operator-O}) has
topological charge $+1$. Hence it has non-vanising matrix elements
between the ground state $|0\rangle$ and one soliton states,
two-particle soliton-breather $sb_n$ states, etc. In this appendix
we determine all one and two-particle form factors. Our results
hold generally for the sine-Gordon model in the attractive regime.

\subsection{One soliton form factors}
The one-soliton form factor is \cite{LukZam01}
\begin{equation}
  F^{1a}_{s}(\theta)=
  \sqrt{Z_1(a)}
  e^{\frac{i\pi a}{2\beta}}
  e^{\frac{a\theta}{\beta}},
  \label{FF-s}
\end{equation}
where
\begin{widetext}
\bea
  \sqrt{Z_1(a)}&=&
  \left(\frac{{\cal C}_2}{2{\cal C}_1^2}\right)^{1/4}
  \left(\frac{\xi{\cal C}_2}{16}\right)^{-1/8}
  \left[
       \frac{\sqrt{\pi}\Delta a_0\Gamma((3+\xi)/2)}
            {v\Gamma(\xi/2)}
  \right]^{2a^2+1/8\beta^2}\nonumber\\
&&\times\  \exp
  \Bigg\{
       \int\limits_{0}^{\infty}\frac{dt}{t}
       \bigg(
       \frac{\cosh(4\xi at/\beta)e^{-(1+\xi)t}-1}
            {4\sinh(t\xi)\sinh(t(\xi+1))\cosh(t)}+
       \frac{1}{1\sinh(t\xi)}-\Bigl(2a^2+\frac{1}{8\beta^2}\Bigr)e^{-2t}
       \bigg)
  \Bigg\}.
  \label{Zn}
\eea
\end{widetext}
\subsection{S-matrices, their analytical properties and minimal
            form factors}
            \label{Subsec-exp-sb-S}

The soliton-breather S-matrix is given by \cite{KarowskiThun}
\begin{widetext}
\bea
  S_{sb_n}(\theta) &=&
  (-1)^n
  \exp
  \Bigg\{
       2\int\limits_{0}^{\infty}\frac{dt}{t}
       \frac{\cosh(\xi t)\sinh(n\xi t)}
            {\cosh(t)\sinh(\xi t)}
       \sinh
       \bigg(
            \frac{2\theta t}{i\pi}
       \bigg)
  \Bigg\}=
  \prod_{j=1}^{n}
  \frac{\sinh\Bigl(\th+i\frac{\pi\xi(n+1-2j)}{2}\Bigr)+
        i\cos\Bigl(\frac{\pi\xi}{2}\Bigr)}
       {\sinh\Bigl(\th-i\frac{\pi\xi(n+1-2j)}{2}\Bigr)-
        i\cos\Bigl(\frac{\pi\xi}{2}\Bigr)}.
  \nn
  \label{S-sbn}
\eea
\end{widetext}
The corresponding soliton-breather 3-particle coupling is
\be
 g_{sb_n}^s=
 \Bigg|
      2
      \cot\bigg(\frac{n\pi\xi}{2}\bigg)
      \prod_{l=1}^{n-1}
      \cot^2\bigg(\frac{l\pi\xi}{2}\bigg)
 \Bigg|^{1/2}.
\ee
Note that by crossing symmetry we must have
\be
  g_{sb_n}^s=g_{\bar{s}s}^{b_n}.
  \label{g-s-sbn}
\ee
The minimal soliton-breather form factor can be obtained combining
equations (2.23), (4.19), and (4.20) of Ref.~\onlinecite{Babuji98},
which can be summarized as
\begin{eqnarray*}
  &&
  S(\theta)=
  \exp
  \Bigg\{
       \int\limits_{0}^{\infty}dt
       f(t)\sinh\bigg(\frac{t\theta}{i\pi}\bigg)
  \Bigg\},
  \\
  &&
\Rightarrow  F^{\rm min}(\theta)=
  \exp
  \Bigg\{
       \int\limits_{0}^{\infty}dt
       f(t)
       \frac{1-\cosh[t(1+\frac{\theta}{i\pi})]}
            {2\sinh(t)}
  \Bigg\}.
\end{eqnarray*}
The soliton-breather S-matrix (\ref{S-sbn}) then gives rise to the
minimal form factor
\begin{eqnarray}
  F^{\rm min}_{sb_n}(\theta) &=&
  {\cal R}_{sb_n}(\theta)
  \exp
  \Bigg\{
       \int\limits_{0}^{\infty}\frac{dt}{t}
       \frac{\cosh(\xi t)\sinh(n\xi t)}
            {\cosh(t)\sinh(\xi t)}
  \times \nonumber \\ && \times
       \frac{1-\cosh[2t(1+\frac{\theta}{i\pi})]}
            {\sinh(2t)}
  \Bigg\},
  \label{F-sbn-min}
\end{eqnarray}
where ${\cal R}_{sb_n}(\theta)$ is given by
\bea
R_{sb_{2n}}(\theta)&=&1\ ,\nonumber\\
R_{sb_{2n+1}}(\theta)&=&i\sinh(\theta/2).
\eea
\subsection{Soliton-breather form factors}
            \label{Subsec-exp-sb-FF}
We will calculate the full form factor using the residue condition
\cite{LukZam01,EsslerKonik}
\begin{eqnarray}
  &&
  ig_{s\bar{s}}^{b_n}F_{sb_n}^{1a}(\theta_s,\theta_b)=
  \nonumber
  \\ && ~~~~~ =
  {\rm Res}_{\delta=0}
  F_{ss\bar{s}}^{1a}
  \Big(
       \theta_s,
       \theta_b-\frac{iu_{b_n}^{\bar{s}s}}{2},
       \theta_b+\delta+\frac{iu_{b_n}^{\bar{s}s}}{2}
  \Big),
  \nonumber
  \\
  \label{Res-FFn}
\end{eqnarray}
where $u_{b_n}^{\bar{s}s}=\pi(1-n\xi)$, and
\begin{equation}
 g_{\bar{s}s}^{b_n}= (-1)^ng_{s\bar{s}}^{b_n}=
 \Bigg|
      2
      \cot\bigg(\frac{n\pi\xi}{2}\bigg)
      \prod_{l=1}^{n-1}
      \cot^2\bigg(\frac{l\pi\xi}{2}\bigg)
 \Bigg|^{1/2}.
 \label{g-bn-ss}
\end{equation}
The three particle form factor involving two solitons and one
antisoliton is \cite{LukZam01}
\begin{widetext}
\begin{eqnarray}
&&F_{ss\bar{s}}^{1a}(\theta_1,\theta_2,\theta_3)=
 \frac{i{\cal C}_2\sqrt{Z_1(a)}}{4{\cal C}_1}
 e^{\frac{i\pi a}{2\beta}}
 e^{\frac{a}{\beta}(\theta_1+\theta_2+\theta_3)+
    \frac{\theta_3}{\xi}}
 G(\theta_{12}) G(\theta_{13}) G(\theta_{23})\nonumber\\
&&\times
 \left[
      e^{\frac{i\pi}{2\beta^2}}
      \int\limits_{C_{+}}\frac{d\th_0}{2\pi}
      e^{-(\frac{2a}{\beta}+\frac{1}{\xi})\th_0}
      W(\theta_{10})W(\theta_{20})W(\theta_{30})
      -e^{\frac{-i\pi}{2\beta^2}}
      \int\limits_{C_{-}}\frac{d\theta_0}{2\pi}
      e^{-(\frac{2a}{\beta}+\frac{1}{\xi})\theta_0}
      W(\theta_{10})W(\theta_{20})W(\theta_{03}) \right],
 \label{FF-2+1}
\end{eqnarray}
where $\theta_{jk}=\theta_j-\theta_k$,
\begin{eqnarray}
  G(\theta)&=&
  i{\cal C}_1
  \sinh\bigg(\frac{\theta}{2}\bigg)
  \exp
  \Bigg\{
      -\int\limits_{0}^{\infty}\frac{dt}{t}
       \frac{\sinh((1-\xi)t)
             \sinh^2(t(1-\frac{i\theta}{\pi}))}
            {\sinh(2t)\cosh(t)\sinh(\xi t)}
  \Bigg\},
  \label{G}\\
  W(\theta)&=&
  \frac{-2}{\cosh(\theta)}
  \exp
  \Bigg\{
       -2
       \int\limits_{0}^{\infty}\frac{dt}{t}
       \frac{\sinh(t(\xi-1))\sinh^2(t(1-i\theta/\pi))}
            {\sinh(2t)\sinh(t\xi)}
  \Bigg\},
  \label{W}
\end{eqnarray}
\end{widetext}
\begin{eqnarray}
  {\cal C}_1=
  \exp
  \left[
      -\int\limits_{0}^{\infty}\frac{dt}{t}
       \frac{\sinh^2(t/2)\sinh(t(\xi-1))}
           {\sinh(2t)\sinh(t\xi)\cosh(t)}
  \right],
  \label{C1}
  \\
  {\cal C}_2=
  \exp
  \left[
       4\int\limits_{0}^{\infty}\frac{dt}{t}
       \frac{\sinh^2(t/2)\sinh(t(\xi-1))}
            {\sinh(2t)\sinh(t\xi)}
  \right].
  \label{C2}
\end{eqnarray}

\begin{figure}[htb]
\centering
\includegraphics[width=70mm,height=73mm,angle=0]{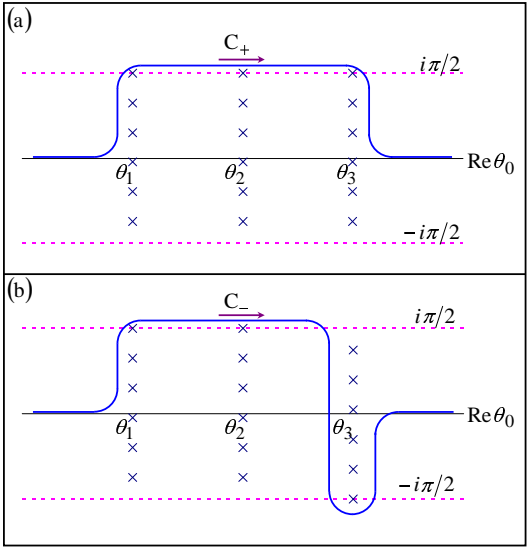}
 \caption{Integration contours $C_{+}$ (panel (a)) and
 $C_{-}$ (panel (b)). Both the contours are located in
 the strip $-\pi/2-0<\Im\theta_0<\pi/2+0$.}
 \label{Fig-Cpm}
\end{figure}

The integration contours $C_{+}$ and $C_{-}$ are constructed as
follows. The contour $C_{+}$ runs from $-\infty$ to $\infty$  in the
complex $\th_0$ plane, passing above the poles at $\theta_p+i\pi/2$,
$p=1,2,3$. Similarly, the contour $C_{-}$ runs above the points
$\theta_p+i\pi/2$, $p=1,2$, and then below $\theta_3-i\pi/2$ (see
figure \ref{Fig-Cpm}).

\begin{figure}[htb]
\centering
\includegraphics[width=70mm,height=37mm,angle=0]{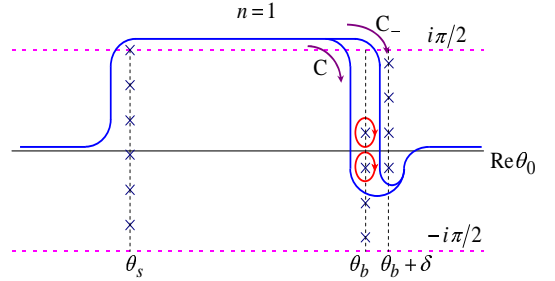}
 \caption{Integration contour $C_{-}$ for $\xi=0.174371$
 and $n=1$. The contour $C_{-}$ is transformed into the contour
 $C$ and the closed contours around the poles.}
 \label{Fig-Cm}
\end{figure}

Let us consider the form factor (\ref{FF-2+1}) for rapidities
$\theta_1=\theta_s$,
$\theta_2=\theta_b-\frac{iu_{b_n}^{\bar{s}s}}{2}$, and
$\theta_3=\theta_b+\delta+\frac{iu_{b_n}^{\bar{s}s}}{2}$. The
function $W(\theta)$ has poles in the strip $|{\rm Im}\
\theta|<\frac{\pi}{2}$ of the complex $\th$-plane at the points
$\theta=iv_k$, where
$$
 v_k=\pi\left(k\xi-\frac{1}{2}\right),
 \ \ \ \ \
 0 \le k < \Big[\frac{1}{\xi}\Big].
$$
As a result the function $W(\theta_{10})W(\theta_{20})W(\theta_{03})$
has poles at $\theta_0=\th_1-i\pi(k\xi-\frac{1}{2})$,
$\th_0=\theta_b+\frac{i\pi\xi}{2}(n-2k)$, and
$\th_0=\theta_b+\delta-\frac{i\pi\xi}{2}(n-2k)$,
$0\le{k}<\frac{n}{2}+\frac{1}{2\xi}$. By construction the contour
$C_{-}$ runs between $n+1$ pairs of poles at
$\th_0=\theta_b+\frac{i\pi\xi}{2}(n-2k)$, and
$\th_0=\theta_b+\delta+\frac{i\pi\xi}{2}(n-2k)$, $0\le{k}\le{n}$,
with only an infinitesimal separation $\delta$ between them (see Fig.
\ref{Fig-Cm}). As a result the integral over $\th_0$ exhibits a simple pole
for $\delta\to 0$. In order to extract the residue of this pole, we deform
the ``singular'' contour $C_{-}$ into a ``regular'' contour $C$ plus
closed contours including one pole from each pair.
$C$ is chosen such that the integral over it is finite in the limit
$\delta\to 0$. The contours $C$ and $C_-$ are shown for $n=1$ in
Fig.~ \ref{Fig-Cm}.

For general $n$ we then find that
\begin{eqnarray}
  &&
  \int\limits_{C_{-}}\frac{d\th_0}{2\pi}
  e^{-(\frac{2a}{\beta}+\frac{1}{\xi})\th_0}
  W(\theta_{10})W(\theta_{20})W(\theta_{03})
  \nonumber \\ && =
  \sum_{k=0}^{n}
  ie^{-(\frac{2a}{\beta}+\frac{1}{\xi})
       (\theta_b+\frac{i\pi\xi(n-2k)}{2})}
  W
  \Big(
      \theta_{1b}-
      \frac{i\pi\xi(n-2k)}{2}
  \Big)
  \nonumber \\ && ~~ \times
  W_{\rm{res}}(iv_k)
  W(-\delta+iv_{n-k})+
  {\rm{regular~part}},
  \label{Cm-Res}
\end{eqnarray}
where $W_{\rm{res}}(iv_k)$ denotes the residue of the function
$W(\theta)$ taken at the point $\theta=iv_k$. We note that
$W(-\delta+iv_{n-k})$ has a simple pole at $\delta=0$.

The regular part of the integral does not contribute to the
residue in the right hand side of equation (\ref{Res-FFn}) and in
the following will be ignored. Similarly, the integral over
$C_{+}$ has a finite limit $\delta\to0$.

In order to proceed we need to analytically continue the function
$W(\theta)$, which can be done using the relation \cite{Babuji98}
\begin{eqnarray}
  W(\theta\pm i\pi\xi)=
  \frac{\sin\left[\frac{i}{2}(\theta\mp\frac{i\pi}{2})\right]}
       {\sin\left[\frac{i}{2}(\theta\pm i\pi\xi\pm\frac{i\pi}{2})\right]}
  W(\theta).
  \label{recurrent}
\end{eqnarray}
Using (\ref{recurrent}) the residues of $W(\theta)$ are readily calculated
\bea
 W_{\rm res}(iv_k)&=&
 \frac{2(-1)^{k-1}}
      {\sqrt{{\cal C}_2}
       \sin(\frac{k\pi\xi}{2})}
 \prod_{j=1}^{k-1}
       \cot
       \left(
            \frac{j\pi\xi}{2}
       \right)\ ,\ k\geq 1,\nonumber\\
  W_{\rm{res}}
  \Big(iv_0  \Big)&=&
 -\frac{2}{\sqrt{{\cal C}_2}}\ .
 \label{Res-W-k}
\end{eqnarray}

The soliton-breather form factors can now be determined from the
residue condition (\ref{Res-FFn})
\begin{eqnarray}
  &&
  ig^{b_n}_{s\bar{s}}F^{1a}_{sb_n}(\theta_s,\theta_b)=
  -\frac{i{\cal{C}}_2\sqrt{Z_1(a)}}
       {4{\cal{C}}_1}
  e^{\frac{i\pi a}{2\beta}+\frac{a\theta_s}{\beta}}
  G(-iu_{b_n}^{\bar{s}s})\nonumber \\ && ~ \times
  G\bigg(\theta_{sb}+\frac{iu_{b_n}^{\bar{s}s}}{2}\bigg)
  G\bigg(\theta_{sb}-\frac{iu_{b_n}^{\bar{s}s}}{2}\bigg)
  \nonumber \\ && ~~ \times
\sum_{k=0}^{n}(-1)^{n-k} e^{-\frac{i\pi a\xi(n-2k)}{\beta}}W_{\rm
  res}(iv_k)W_{\rm res}(iv_{n-k})\nonumber\\
&&\qquad\qquad\times \ W\Bigl(\theta_{sb}-i\frac{\pi\xi(n-2k)}{2}\Bigr).
  \label{FF-s-b2m-1}
\end{eqnarray}

The soliton-breather form factor can be expressed in terms of the
minimal form factor (\ref{F-sbn-min}) as
\begin{eqnarray}
  F^{1a}_{sb_n}(\theta_s,\theta_b) &=&
  N^{a}_{sb_n}
  \sqrt{Z_1(a)}
  e^{\frac{i\pi a}{2\beta}+\frac{a\theta_s}{\beta}}
  K^{a}_{sb_n}(\theta_{sb})
  \times \nonumber \\ && \times
  F_{sb_n}^{\rm min}(\theta_{sb}).
  \label{ff-s-b-2m}
\end{eqnarray}
Here $N^{a}_{sb_n}$ is a normalization constant
given by
\begin{eqnarray}
  N^{a}_{sb_n}=
  \frac{e^{\frac{i\pi n\xi a}{\beta}}}
       {ig^{s}_{sb_n}
        {\rm{Res}}_{\theta=iu^{sb_n}_{s}}
        \left[
            K^{a}_{sb_n}(\theta)
        \right]
        F_{sb_n}^{\rm{min}}(iu^{sb_n}_{s})},
  \label{N-sbn}
\end{eqnarray}
where
\be
u_s^{sb_k}=\frac{\pi}{2}(k\xi+1).
\ee
The ``pole functions'' $K^{a}_{sb_n}(\theta)$ have somewhat different
forms for even and odd $n$ respectively and are given by
\begin{widetext}
\begin{eqnarray}
  K^{a}_{sb_{2m}}(\theta)&=&\frac{1}{\cosh\theta}
  \prod_{l=1}^{m-1}
\frac{1}
       {\cosh\Bigl(\frac{\theta}{2}+\frac{iu_s^{sb_{2l}}}{2}\Bigr)
        \cosh\Bigl(\frac{\theta}{2}-\frac{iu_s^{sb_{2l}}}{2}\Bigr)}
\nonumber\\
&&\times\quad
  \bigg\{
       \Big[
           W_{\rm res}(iv_m)
       \Big]^2-
       \sum_{k=1}^{m}
       W_{\rm res}(iv_{m-k})
       W_{\rm res}(iv_{m+k})
       \Big[
           e^{-\frac{2i\pi ka\xi}{\beta}}
           \Psi_{2k}(\theta)+
           e^{\frac{2i\pi ka\xi}{\beta}}
           \Psi_{2k}(-\theta)
       \Big]
  \bigg\},
  \label{K-2m}\\
  K^{a}_{sb_{2m+1}}(\theta)&=&
  \frac{1}{\sinh(\frac{\theta}{2})}
  \prod_{l=0}^{m-1}
\frac{1}
       {\cosh\Bigl(\frac{\theta}{2}+\frac{iu_s^{sb_{2l+1}}}{2}\Bigr)
        \cosh\Bigl(\frac{\theta}{2}-\frac{iu_s^{sb_{2l+1}}}{2}\Bigr)}
\nonumber \\ && \times
  \sum_{k=0}^{m}
  W_{\rm res}(iv_{m-k})
  W_{\rm res}(iv_{m+k+1})
  \left[
       e^{-\frac{i\pi a(2k+1)\xi}{\beta}}
       \Psi_{2k+1}(\theta)+
       e^{\frac{i\pi a(2k+1)\xi}{\beta}}
       \Psi_{2k+1}(-\theta)
\right],
  \label{K-2m+1}
\end{eqnarray}
where
\end{widetext}
\be
 \Psi_{2k}(\theta) =
 \frac{\sin(\frac{i\theta}{2}-\frac{\pi}{4})}
      {\sin\Bigl(\frac{i\theta}{2}+\frac{u_s^{sb_{2k}}}{2}\Bigr)}
 \prod_{j=1}^{k-1}
 \cot
 \Bigl(\frac{i\theta}{2}+\frac{u_s^{sb_{2j}}}{2} \Bigr),
 \label{Psi-2k}
\ee
\be
  \Psi_{2k+1}(\theta) =
  \frac{1}
       {\sin\Bigl(\frac{i\theta}{2}+\frac{u_s^{sb_{2k+1}}}{2}\Bigr)}
  \prod_{j=0}^{k-1}
  \cot
  \Bigl(
      \frac{i\theta}{2}+\frac{u_s^{sb_{2j+1}}}{2}
  \Bigr).
  \label{Psi-2k+1}
\ee

\section{Form factors of $\exp[ia\Theta]$}
        \label{Sec-exp-FF-bb}
The operator $\exp[ia\Theta]$ carries zero topological
charge. Hence the non-vanishing form factors with less than three
particles involve a single breather, a soliton-antisoliton pair or two
breathers. In this appendix we construct these one and two particle form
factors.

\subsection{S-matrices and minimal form factors}
            \label{Subsec-sa-Smatr}
The breather-breather S-matrix is \cite{KarowskiThun}
\begin{eqnarray}
  &&
  S_{b_kb_l}(\theta)=
  \exp
  \Bigg\{
      4\int\limits_{0}^{\infty}\frac{dt}{t}
      \frac{\sinh(\frac{2\theta t}{i\pi})}
           {\cosh(t)\sinh(\xi t)}
  \times \label{S-kl-exp} \\ && ~~~~~ \times
  \cosh(\xi t)\sinh(k\xi t)\cosh((1-l\xi)t)
  \Bigg\},
  \ \ \ k<l,
  \nonumber
  \\
  &&
  S_{b_kb_k}(\theta)=
  -\exp
  \Bigg\{
      2\int\limits_{0}^{\infty}\frac{dt}{t}
      \frac{\sinh(\frac{2\theta t}{i\pi})}
           {\cosh(t)\sinh(\xi t)}
  \times \label{S-kk-exp} \\ && ~~~~~ \times
  \Big[
      \cosh(\xi t)\sinh((2k\xi-1)t)+\sinh((1-\xi)t)
  \Big]
  \Bigg\}.
  \nonumber
\end{eqnarray}
Evaluating the integrals gives
\begin{eqnarray}
  S_{b_kb_l}(\theta) &=&
  \tanh
  \bigg(
       \frac{\theta+iu^{bb}_{l+k}}{2}
  \bigg)
  \coth
  \bigg(
       \frac{\theta-iu^{bb}_{l+k}}{2}
  \bigg)
  \times\nonumber\\&\times&
  \tanh
  \bigg(
       \frac{\theta+iu^{bb}_{l-k}}{2}
  \bigg)
  \coth
  \bigg(
       \frac{\theta-iu^{bb}_{l-k}}{2}
  \bigg)
  \times \nonumber
  \\&\times&
  \prod_{j=1}^{k-1}
  \Bigg\{
       \tanh^2
       \bigg(
            \frac{\theta+iu^{bb}_{l-k+2j}}{2}
       \bigg)
  \times \label{S-kl-TgCtg} \\ && ~~~~~ \times
       \coth^2
       \bigg(
            \frac{\theta-iu^{bb}_{l-k+2j}}{2}
       \bigg)
  \Bigg\},
  \nonumber
\end{eqnarray}
where now ${k}\le{l}$, and \be
 u^{bb}_{n}=\frac{n\pi\xi}{2}.
\ee The simple pole in $S_{b_kb_l}(\theta)$ at $\theta=i
u_{k+l}^{bb}$ corresponds to the formation of a $b_{k+l}$ breather
bound state. The corresponding residue is
\begin{eqnarray}
S_{\rm{res}}(iu^{bb}_{k+l}) &=&
2i\tan\Big(\frac{(k+l)\pi\xi}{2}\Big)
  \tan
  \Big(\frac{l\pi\xi}{2}\Big)
  \cot
  \Big(\frac{k\pi\xi}{2}\Big)
  \nonumber \\ &\times&
  \prod_{j=1}^{k-1}
  \cot^2
  \bigg(
       \frac{j\pi\xi}{2}
  \bigg)
  \prod_{j=l+1}^{l+k-1}
  \tan^2
  \bigg(
       \frac{j\pi\xi}{2}
  \bigg).
\end{eqnarray}
The corresponding three-particle coupling is
\begin{eqnarray}
  g^{b_{k+l}}_{b_kb_l} &=&
        \bigg|
             2
             \cot\bigg(\frac{k\pi\xi}{2}\bigg)
             \tan\bigg(\frac{l\pi\xi}{2}\bigg)
             \tan\bigg(\frac{(k+l)\pi\xi}{2}\bigg)
        \bigg|^{\frac{1}{2}}
  \nonumber\\ & \times&
  \prod_{j=1}^{k-1}
  \Bigg|
      \cot
      \bigg(
          \frac{j\pi\xi}{2}
      \bigg)
      \tan
      \bigg(
          \frac{(l+j)\pi\xi}{2}
      \bigg)
  \Bigg|.
\end{eqnarray}
The minimal form factor for two different breathers is then
\begin{eqnarray}
  F_{b_kb_l}^{\rm min}(\theta) &=&
  \exp
  \Bigg\{
      -2\int\limits_{0}^{\infty}\frac{dt}{t}~
       \frac{\cosh(\xi t)\cosh((1-l\xi)t)}
            {\cosh(t)}
  \times \nonumber \\ && \times
       \frac{\sinh(k\xi t)}{\sinh(\xi t)}~
       \frac{\cosh(2t(1-\frac{i\theta}{\pi}))-1}
            {\sinh(2t)}
  \Bigg\}.
  \label{FF-kl-min}
\end{eqnarray}

The minimal form factor $F_{b_kb_k}^{\rm min}(\theta)$ involving two
breathers of the same type is given by
\begin{eqnarray}
  &&
  F_{b_kb_k}^{\rm min}(\theta)=
  i\sinh\bigg(\frac{\theta}{2}\bigg)
  \exp
  \Bigg\{-2
       \int\limits_{0}^{\infty}\frac{dt}{t}
       \frac{\sinh^2(t(1-\frac{i\theta}{\pi}))}
            {\sinh(2t)}
  \nonumber \\ && ~~~~~ \times
       \frac{\cosh(\xi t)\sinh((2k\xi-1)t)+
             \sinh((1-\xi)t)}
            {\cosh(t)\sinh(\xi t)}
  \Bigg\}.
  \nonumber \\
  \label{FF-kk-min-1}
\end{eqnarray}

\subsection{Soliton-antisoliton and one-breather form factors}
            \label{Subsec-sa-exp}
The soliton-antisoliton form factors for the operators
$\cos(\beta\Theta)$, $\sin(\beta\Theta)$, and $\Theta$,
\begin{equation}
  \Theta=
  -i\lim_{a\to0}\partial_{a}e^{ia\Theta},
  \label{theta-exp}
\end{equation}
are \cite{Luk}
\bea
  &&
  F^{\cos(\beta\Theta)}_{s\bar{s}}(\theta)=
  \frac{{\cal G}_{\beta}}{2}
  \frac{G(\theta)}{{\cal C}_1}
  \cot
  \bigg(
      \frac{\pi\xi}{2}
  \bigg)
  \frac{8i\cosh(\frac{\theta}{2})}
       {\xi\sinh(\frac{\theta-i\pi}{\xi})}
  \times \nonumber \\ && ~~~~~~~~~~~~~~~ \times
  \cosh
  \bigg(
      \frac{\theta-i\pi}{2\xi}
  \bigg),
  \label{FF-sa-cos}
\eea
\bea  &&
  F^{\sin(\beta\Theta)}_{s\bar{s}}(\theta)=
 -\frac{{\cal G}_{\beta}}{2}
  \frac{G(\theta)}{{\cal C}_1}
  \cot
  \bigg(
      \frac{\pi\xi}{2}
  \bigg)
  \frac{8\cosh(\frac{\theta}{2})}
       {\xi\sinh(\frac{\theta-i\pi}{\xi})}
  \times \nonumber \\ && ~~~~~~~~~~~~~~~ \times
  \sinh
  \bigg(
      \frac{\theta-i\pi}{2\xi}
  \bigg),
  \label{FF-sa-sin}
\eea
\be
  F^{\Theta}_{s\bar{s}}(\theta)=
  -\frac{G(\theta)}{{\cal C}_1}
  \frac{\pi}
       {\beta
        \cosh(\frac{\theta-i\pi}{2\xi})
        \cosh(\frac{\theta}{2})},
  \label{FF-sa-Phi}
\ee
where $G(\theta)$ and ${\cal{C}}_1$ are given by eqs. (\ref{G}) and
(\ref{C1}), respectively, and
\begin{widetext}
\begin{eqnarray}
{\cal G}_a \equiv \langle e^{ia\Theta}\rangle
=\Bigg[
       \frac{a_0\Delta\sqrt{\pi}\Gamma(\frac{1}{2-2\beta^2})}
            {v2\Gamma(\frac{\beta^2}{2-2\beta^2})}
  \Bigg]^{2a^2}\exp \Bigg\{
   \int\limits_{0}^{\infty}\frac{dt}{t}\bigg[-2a^2e^{-2t}
+\frac{\sinh^2(2a\beta
       t)}{2\sinh(t\beta^2)\sinh(t)\cosh(t(1-\beta^2))} \bigg]
  \Bigg\}.&&
\end{eqnarray}
\end{widetext}
The single-particle form factors $F^{\beta}_{b_n}$ and
$F^{\Theta}_{b_n}$ for the operators $e^{i\beta\Theta}$ and
$\Theta$, respectively, can be obtained from the residue condition for
the soliton-antisoliton form factor $F^{\beta}_{s\bar{s}}(\theta)$
and $F^{\Theta}_{s\bar{s}}(\theta)$
$$
 g^{b_n}_{s\bar{s}}F^{a}_{b_n}=
 {\rm Res}_{\theta=-iu^{s\bar{s}}_{b_n}}F^{a}_{s\bar{s}}(\theta),
 \ \ \
 u^{s\bar{s}}_{b_n}=\pi(1-n\xi),
$$
where $a=\beta,\Theta$. Using equations (\ref{FF-sa-cos}),
(\ref{FF-sa-sin}), and (\ref{FF-sa-Phi}), we can write
\begin{widetext}
\begin{eqnarray}
  F^{\beta}_{b_n}&=&\frac{2\cot\Bigl(\frac{\pi\xi}{2}\Bigr)\sin(n\pi\xi)
(-i)^n}{\left[2\cot\Bigl(\frac{\pi\xi n}{2}\Bigr)
\prod_{l=1}^{n-1}\cot^2\Bigl(\frac{\pi j\xi}{2}\Bigr)
\right]^{\frac{1}{2}}}
{\cal G}_\beta
\exp\Bigl[\int_0^\infty\frac{dt}{t}
\frac{\sinh( t(\xi-1))\sinh^2(tn\xi)}{\sinh(2t)\cosh(t)\sinh(t\xi)}
\Bigr],
  \label{FF-bn-beta}\\
F^{\Theta}_{b_{2m-1}}&=&
  i(-1)^{m-1}
  \frac{G(i\pi((2m-1)\xi-1))}
       {g^{b_{2m-1}}_{s\bar{s}}{\cal C}_1}
  \frac{\pi\xi}
       {\beta\sin\left(\frac{(2m-1)\pi\xi}{2}\right)}.
  \label{FF-Phi-bn}
\end{eqnarray}
\end{widetext}
\subsection{Breather-breather form factors of $\exp(ia\Theta)$}
           \label{Subsec-bkbl-exp}
To calculate two breather form factor $F^a_{b_kb_l}(\theta_{12})$,
we will start with the formulas for $n$-breather ($n=k+l$) form
factor $F^a_{\underline{b_1}}(\underline{\gamma})$ from
Ref.\cite{BabKar02},
\begin{eqnarray}
  &&
  F^a_{\underline{b_1}}(\underline{\gamma})=
  \frac{{\cal G}_a}
       {2^{n/2}}
  K^a_n(\underline{\gamma})
  \prod_{1 \le i < j \le n}
  R(\gamma_{ij}),
  \label{FF-n-b1-breathers}
\end{eqnarray}
\begin{eqnarray}
 &&
  R(\gamma)=
  \frac{{\cal N}F_{b_1b_1}^{min}(\gamma)}
       {\sinh(\frac{1}{2}(\gamma-i\pi\xi))
        \sinh(\frac{1}{2}(\gamma+i\pi\xi))},
  \label{R}
  \\
  &&
  K^a_n(\underline{\gamma})=
  \sum_{l_1=0}^{1} \cdots \sum_{l_n=0}^{1}
  (-1)^{l_1+ \cdots +l_n}
  p^a_n(\underline{l})
  \times \nonumber \\ && ~~~~~~~~~~ \times
  \prod_{1 \le i < j \le n}
  \bigg[
       1+(l_i-l_j)\frac{i\sin(\pi\xi)}{\sinh(\gamma_{ij})}
  \bigg],
  \label{Kn}
  \\
  &&
  p^a_n(\underline{l})=
  \bigg(
       \frac{2}{R(-i\pi)\sin(\pi\xi)}
  \bigg)^{n/2}
  \prod_{i=1}^{n}
  e^{\frac{i\pi\xi a}{\beta}(-1)^{l_i}},
  \nonumber \\
  \label{pn}
\end{eqnarray}
where $\underline{\gamma}=(\gamma_1, \ldots, \gamma_n),$
$\underline{l}=(l_1, \ldots, l_n),$
$\gamma_{ij}=\gamma_{i}-\gamma_{j}$,
\begin{eqnarray*}
  {\cal N} &=&
  -\exp
  \Bigg\{
      2\int\limits_{0}^{\infty}\frac{dt}{t}
       \frac{\sinh(\xi t)\sinh((1-\xi)t)}
            {\cosh(t)\sinh(2t)}
  \Bigg\}.
\end{eqnarray*}
The normalization factor ${\cal G}_a/2^{n/2}$ is introduce in
Ref.\cite{Luk}

Form factor $F_{b_kb_l}(\theta_1,\theta_2)$ can be calculated as
follows.
\begin{itemize}
\item First step, we calculate the residues of the form factor
      at the point $\gamma_{n-1}-\gamma_n=-i\pi\xi$
      ($\gamma_{n-1}=\gamma'_{n-1}-\frac{i\pi\xi}{2}$,
      $\gamma_n=\gamma'_{n-1}+\delta_1+\frac{i\pi\xi}{2}$).
      As a result, we obtain the form factor for $n-2$
      breathers $b_1$ and one breather $b_2$.
\item Second step, we calculate the residue of the $k+l-1$-particle form
      factor obtained at the point
      $\gamma_{n-2}-\gamma'_{n-1}=\frac{3i\pi\xi}{2}$
      ($\gamma_{n-2}=\gamma'_{n-2}-i\pi\xi$,
      $\gamma'_{n-1}=\gamma'_{n-2}+\delta_2+\frac{i\pi\xi}{2}$).
      As a result, we obtain the form factor for $n-3$
      breathers $b_1$ and one breather $b_3$.
\item Step number $l-1$,  we calculate the residue of the $k+2$-particle form
      factor obtained at the point $\gamma_{k+1}-\gamma'_{k+2}=\frac{ki\pi\xi}{2}$
      ($\gamma_{k+1}=\gamma'_{k+1}-\frac{i(l-1)\pi\xi}{2}$,
      $\gamma'_{k+2}=\gamma'_{k+1}+\delta_{l-1}+\frac{i\pi\xi}{2}$).
      As a result, we obtain the form factor for $k$
      breathers $b_1$ and one breather $b_l$.
\end{itemize}
Finally, taking $\gamma'_{k+1}\equiv\theta_2$, we can write
\begin{eqnarray}
  \gamma_{k+m} &=&
  \theta_2+\sum_{j=l-m+1}^{l-1}\delta_j-
  \frac{i(l-2m+1)\pi\xi}{2},
  \label{gamma-theta1}
\end{eqnarray}
$m=1,2,\ldots,l$, and $\delta_j$ are infinitesimal parameters.

Similar calculations performed with the variables $\gamma_j$,
$j=1,2,\ldots,k$, give
\begin{eqnarray}
  \gamma_{m} &=&
  \theta_1+\sum_{j=k-m+1}^{k-1}\epsilon_j-
  \frac{i(k-2m+1)\pi\xi}{2},
  \label{gamma-theta2}
\end{eqnarray}
$m=1,2,\ldots,k$, and $\epsilon_j$ are infinitesimal parameters.

It should be noted that the order of calculation of residues
predicts the rules
\begin{eqnarray}
 &&
 \begin{array}{ccccccccc}
        |\delta_1| & \ll &
        |\delta_2| & \ll &
        \cdots & \ll &
        |\delta_{k-1}| & \ll &
        |\theta_{12}|,
        \\
        |\epsilon_1| & \ll &
        |\epsilon_2| & \ll &
        \cdots & \ll &
        |\epsilon_{l-1}| & \ll &
        |\theta_{12}|.
 \end{array}
 \label{rule-k+l}
\end{eqnarray}

The $n$-particle ($n=k+l$) form factor (\ref{FF-n-b1-breathers})
depends on $\gamma_i-\gamma_j$, $1 \le i \le j \le n$. Taking into
account equations (\ref{gamma-theta2}), (\ref{gamma-theta1}), and
(\ref{rule-k+l}), we can write
\begin{eqnarray}
 &&
 \gamma_{ij}=
 \left\{
      \begin{array}{l}
            -\epsilon_{k-j}-
             i(j-i)\pi\xi,
             {\rm ~~if~} (i,j) \in A_1,
            \\
            -\delta_{l-j}-
             i(j-i)\pi\xi,
             {\rm ~~if~} (i,j) \in A_2,
            \\
             \theta_{12}-
             i
             \Big(
                 \frac{l-k}{2}+j-m
             \Big),
             {\rm ~~if~} (i,j) \in A_3,
      \end{array}
 \right.
 \label{gamma-ij-theta-12}
\end{eqnarray}
where the manifolds $A_{1,2,3}$ are  constructed as following,
\begin{eqnarray*}
  &&
  A_1: ~ 1 \le i \le j \le k,
  \\
  &&
  A_2: ~ k+1 \le i \le j \le n,
  \\
  &&
  A_3: ~ 1 \le i \le k, ~ k+1 \le j \le n.
\end{eqnarray*}
Then we obtain the following expression for the two-particle form
factor,
\begin{eqnarray}
  F^a_{b_kb_l}(\theta_{12}) &=&
  N^a_{b_kb_l}
  K^a_{b_kb_l}(\theta_{12})
  F_{b_kb_l}^{\rm{min}}(\theta_{12}),
  \label{FF-beta-kl}
\end{eqnarray}
where the minimal form factor $F_{b_kb_l}^{\rm{min}}(\theta_{12})$
is given by equation (\ref{FF-kl-min}) for $k<l$, and equation
(\ref{FF-kk-min-1}) for $k=l$,
\begin{eqnarray}
  &&
  K^a_{b_kb_l}(\theta)=
  K^a_{n}(\underline{\gamma})
  \times \nonumber \\ && ~~~~~ \times
  \prod_{\nu=\frac{l-k}{2}+1}^{\frac{l+k}{2}}
  \frac{1}
       {\sinh
        \big(
             \frac{1}{2}
             (\theta-i\pi\xi\nu)
        \big)
        \sinh
        \big(
             \frac{1}{2}
             (\theta+i\pi\xi\nu)
        \big)}.
  \nonumber \\
  \label{K-a-kl}
\end{eqnarray}
The normalization constant $N^{a}_{b_kb_l}$ can be calculated as
\begin{eqnarray*}
  &&
  N^a_{b_kb_l}=
  \frac{ig^{k+l}_{kl}F^a_{b_{k+l}}}
       {K_{b_kb_l}^{\rm{res}}
        F_{b_kb_l}^{\rm{min}}(-\frac{i\pi\xi}{2}(k+l))},
  \\
  &&
  K_{b_kb_l}^{\rm{res}}=
  {\rm{Res}}_{\delta=0}[K^a_{b_kb_l}(-\delta-\frac{i\pi\xi}{2}(k+l))],
\end{eqnarray*}
the one-particle form factor $F_{b_n}$ is given by equation
(\ref{FF-bn-beta}).

In particular, the pole function for the few lowest breathers are
\begin{widetext}
\begin{eqnarray*}
  K^a_{b_1b_1}(\theta) &=&
  \frac{[a]^2}
       {\sinh(\frac{1}{2}
              (\theta-i\pi\xi))
        \sinh(\frac{1}{2}
              (\theta+i\pi\xi))},
  \\
  K^a_{b_1b_2}(\theta) &=&
  \frac{[a]}
       {\sinh(\frac{1}{2}(\theta-\frac{3i\pi\xi}{2}))
        \sinh(\frac{1}{2}(\theta+\frac{3i\pi\xi}{2}))}
  \nonumber
  \bigg\{
       [a]^2+
       \frac{1}
            {8
             \cos\big(\frac{\pi\xi}{2}\big)
             \cosh\big[\frac{1}{2}\big(\theta+\frac{i\pi\xi}{2}\big)\big]
             \cosh\big[\frac{1}{2}\big(\theta-\frac{i\pi\xi}{2}\big)\big]}
  \bigg\},
  \\
  K^a_{b_2b_2}(\theta) &=&
  \prod_{k=1}^{2}
  \frac{1}
       {\sinh(\frac{1}{2}(\theta-ik\pi\xi))
        \sinh(\frac{1}{2}(\theta+ik\pi\xi))}
  [a]^2
  \bigg\{
       [a]^2+
       \frac{1}
            {2
             \cosh\big[\frac{1}{2}\big(\theta+i\pi\xi\big)\big]
             \cosh\big[\frac{1}{2}\big(\theta-i\pi\xi\big)\big]}
  \bigg\},
  \\
  K^a_{b_1b_3}(\theta) &=&
  \frac{[a]^2}
       {\sinh(\frac{1}{2}(\theta-2i\pi\xi))
        \sinh(\frac{1}{2}(\theta+2i\pi\xi))}
  \Bigg\{
       [a]^2+
       \frac{1}
            {4\cos(\pi\xi)\big(1+\cos(\pi\xi)\big)}+
       \\ && +
       \frac{1+2\cos(\pi\xi)}
            {8
             \cos(\pi\xi)
             \cosh
             \left(
                  \frac{1}{2}
                  \big(
                       \theta+
                       i\pi\xi
                  \big)
             \right)
             \cosh
             \Big(
                  \frac{1}{2}
                  \big(
                      \theta-
                      i\pi\xi
                  \big)
             \Big)}
  \Bigg\}.
\end{eqnarray*}
\end{widetext}
where
$$
 [a]=\frac{\sin(\frac{\pi\xi a}{\beta})}
          {\sin(\pi\xi)}.
$$

\subsection{Breather-breather form factors of $\Theta$}
           \label{Subsec-bkbl-Phi}

Equation (\ref{theta-exp}) allows us to express the operator
$\Theta$ in terms of exponential operator $e^{ia\Theta}$.
Therefore the form factors $F^{\Theta}_{b_kb_l}(\theta)$ are
expressed in terms of the form factors of the exponential fields.
Taking into account that the pole function
$K^{a}_{b_kb_l}(\theta)$ (\ref{K-a-kl}) goes to zero as $a^2$ when
$k+l$ is even and linearly with $a$ when $k+l$ is odd, we can
conclude that only form factors $F^{\Theta}_{b_kb_l}(\theta)$ with
$k+l$ being odd are nontrivial. These form factors are written as
\begin{eqnarray}
  F^{\Theta}_{b_kb_l}(\theta) &=&
  N^{\Theta}_{b_kb_l}
  K^{\Theta}_{b_kb_l}(\theta)
  F^{\rm{min}}_{b_kb_l}(\theta),
  \label{FF-Phi-kl}
\end{eqnarray}
where the pole function $K^{\Theta}_{b_kb_l}(\theta)$ is given by,
$$
 K^{\Theta}_{b_kb_l}(\theta)=
 -i\lim_{a\to0}
 \partial_a
 K^{a}_{b_kb_l}(\theta),
$$
the normalization factor $N^{\Theta}_{b_kb_l}$ is determined as
$$
 N^{\Theta}_{b_kb_l}=
 \lim_{a\to0}
 N^{a}_{b_kb_l},
$$
and the minimal form factor is defined by equation
(\ref{FF-kl-min}).

\end{document}